\documentclass{article}

\usepackage[preprint]{neurips_2026}

\usepackage{mathtools}
\usepackage[utf8]{inputenc} % allow utf-8 input
\usepackage[T1]{fontenc}    % use 8-bit T1 fonts
\usepackage{hyperref}       % hyperlinks
\usepackage{url}            % simple URL typesetting
\usepackage{booktabs}       % professional-quality tables
\usepackage{amsfonts}       % blackboard math symbols
\usepackage{nicefrac}       % compact symbols for 1/2, etc.
\usepackage{microtype}      % microtypography
\usepackage{xcolor}         % colors
\usepackage{graphicx}
\usepackage{subcaption}
\usepackage{booktabs} % for professional tables
\usepackage{bbm}
\usepackage{bm}
\usepackage{amsmath}
\usepackage{comment}

\newtheorem{theorem}{Result}
\newtheorem{theorem_proper}{Theorem}
\newtheorem{corollary}{Corollary}

\def\x{\bm x}
\def\w{\bm w}

\title{Theory of Optimal Learning Rate Schedules and Scaling Laws for a Random Feature Model}

\author{%
Blake Bordelon$^{1,2}$\thanks{Equal contribution} \ , \ 
Francesco Mori$^{1}$\footnotemark[1]
\\
$^{1}$Center for Mathematical Sciences and Applications, Harvard University\\
$^{2}$Oden Institute for Computational Engineering and Sciences \& Dept. of Neuroscience, UT Austin\\
\texttt{\{  blake , francesco \}@cmsa.fas.harvard.edu}
}

\begin{document}

\maketitle

\begin{abstract}
  Setting the learning rate (LR) for a deep learning model is a critical part of successful training. Choosing LRs is often done empirically with trial and error. In this work, we explore a solvable model of optimal LR schedules for a powerlaw random feature model trained with stochastic gradient descent (SGD). We consider the optimal schedule $\eta_T^\star(t)$ where $t$ is the current iterate and $T$ is the training horizon. This schedule is computed both as a numerical optimization problem and also analytically using optimal control theory. Our analysis reveals two regimes which we term the easy phase and hard phase. In the easy phase the optimal schedule is a polynomial decay $\eta_T^\star(t) \simeq T^{-\xi} (1-t/T)^{\delta}$ where $\xi$ and $\delta$ depend on the properties of the features and task. In the hard phase, the optimal schedule resembles warmup-stable-decay with constant initial LR and annealing performed over a vanishing fraction of training steps. We investigate joint optimization of LR and batch size and find batch ramps can improve the wall-clock time in the easy phase. Beyond SGD, we derive optimal schedules for momentum parameter $\beta(t)$ and show that it improves the loss-scaling exponent in the hard phase. We compare our optimal schedule to various benchmarks including (1) optimal constant learning rates $\eta_T(t) \sim T^{-\xi}$ (2) optimal power laws $\eta_T(t) \sim  T^{-\xi} t^{-\chi}$, finding that our schedule achieves better rates than either of these. Our theory suggests that LR transfer across training horizon depends on the structure of the model and task. For ResNet image classification on CIFAR-5M, the learning curves exhibit hard-phase behavior where optimal base LRs are constant under sufficient annealing. GPT-2 style transformers trained in language modeling exhibit easy-phase behavior where optimal LRs shift even under annealing. 
\end{abstract}

\section{Introduction}

\begin{figure*}
    \centering
\begin{subfigure}[b]{0.37\linewidth}
        \centering
        % Replace with your actual filename for easy phase schedule
       \includegraphics[width=\linewidth]{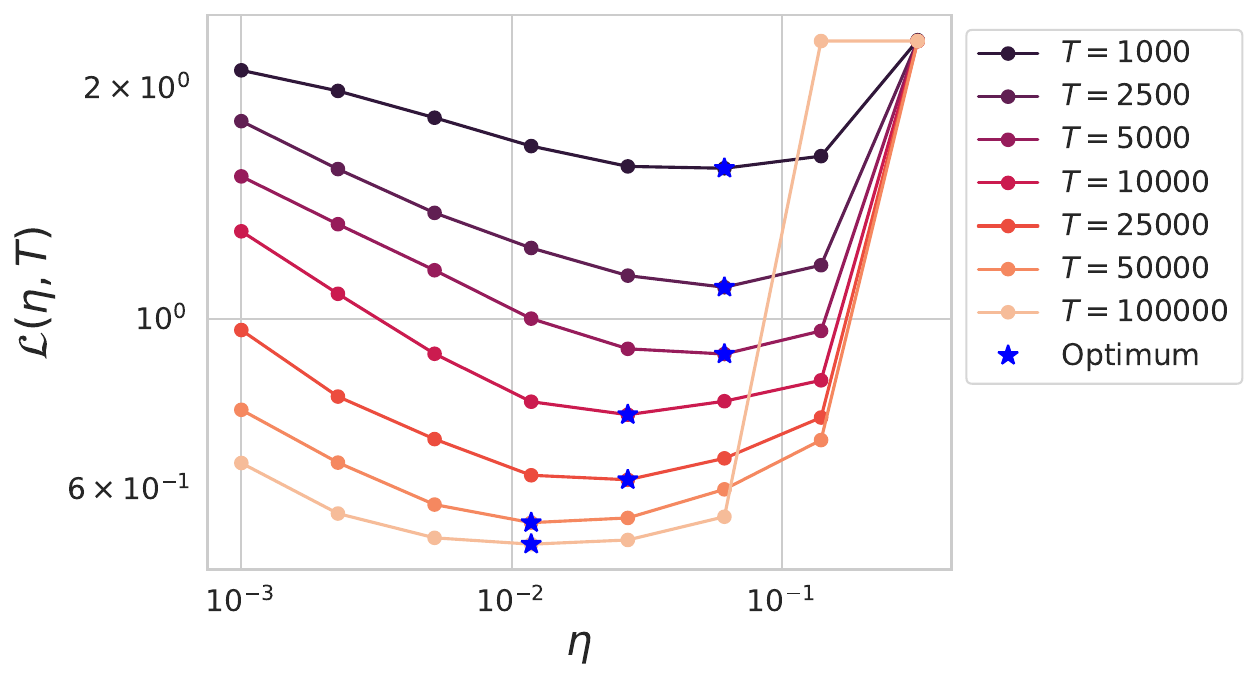}
        \caption{ResNet on CIFAR-5M}
    \end{subfigure}
    \begin{subfigure}[b]{0.265\linewidth}
            \includegraphics[width=\linewidth]{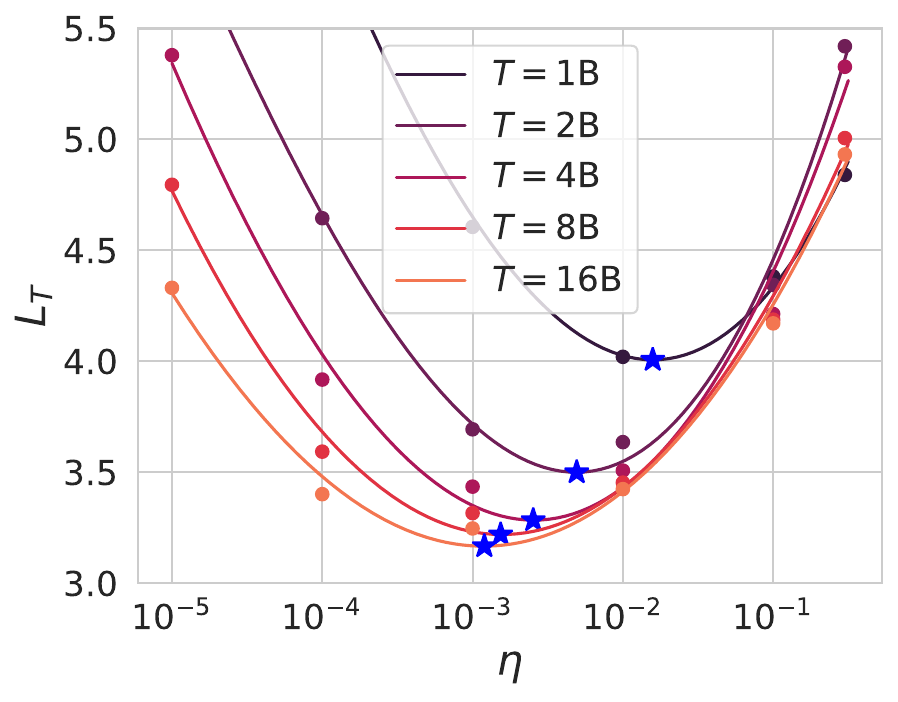}
            \caption{Transformer on C4}
    \end{subfigure}
    \begin{subfigure}[b]{0.35\linewidth}
            \includegraphics[width=\linewidth]{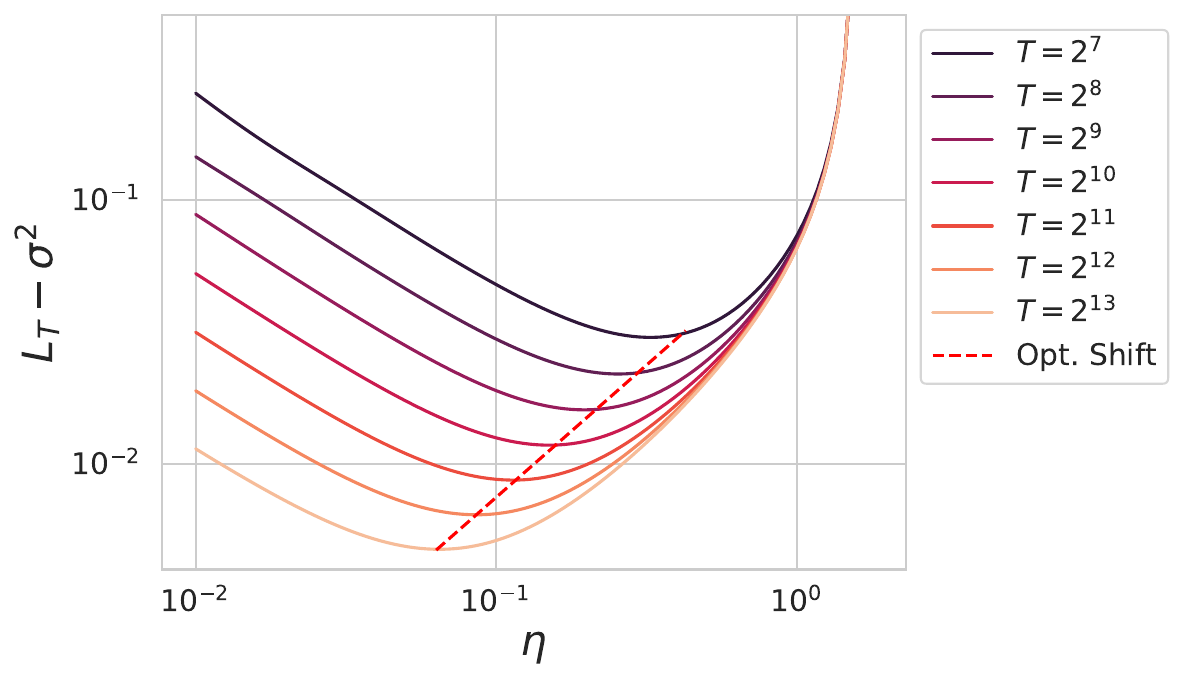}
            \caption{Random Feature Model}
    \end{subfigure}
    
    \caption{Learning rates do not automatically transfer over training horizons $T$. This motivates theory that can identify not only how to scale $\eta$ with $T$, but also how to set the entire learning rate schedule $\eta(t)$ with $T$. (a) The loss of a deep ResNet trained on CIFAR-5M. (b) GPT-style transformer trained on next word prediction on C4 dataset. (c) Test loss of a random feature model trained with SGD as a function of fixed learning rate $\eta$.  }
    \label{fig:token_horizon_transfer_failure}
\end{figure*}

Training deep learning models requires choosing many hyperparameters such as the learning rate (LR), batch size, training horizon, total data, and architectural details, resulting in a complicated decision space. To make matters worse, it is challenging to characterize how these various hyperparameters jointly interact as one scales up the model size or training horizon. One strategy to reduce this complexity is to identify scaling protocols that allow for \textit{transfer} of optimal hyperparameters, where small models can provide reliable proxies for tuning for large models. Such approximate hyperparameter transfer across model sizes can be achieved by principled parameterization and optimizer design \cite{yang2022tensor, bordelon2023depthwise,yang2023feature, dey2025don}. However, these schemes do not automatically lead to transfer over training horizons \cite{bjorck2024scaling, everett2024scaling}. In Figure \ref{fig:token_horizon_transfer_failure} we illustrate the failure of transfer over training horizons $T$ despite having successful transfer over model sizes (widths $N$). This failure motivates theory that can account for the behavior of optimal learning rates of SGD as the training horizon varies. Further, beyond optimal base learning rates themselves, one can also adopt a \textbf{learning rate schedule}. While many schedules such as linear decay, warmup-stable-decay, and cosine annealing are commonly used in practice, it is unclear under which conditions one should favor one type of schedule over the other.

To address these questions in a theoretical framework, we make the following contributions:
\begin{itemize}
    \item We study SGD dynamics in a powerlaw random feature model \cite{bordelon2022learning, bordelon2024a, paquette2024}. We show SGD fluctuations can cause shifts in optimal learning rates across training horizons $T$ that are \textbf{task and model dependent}. 
    \item We characterize both numerically and analytically the \textbf{optimal learning rate schedule} for SGD in our model using optimal control, minimizing the final test loss at step $T$. This analysis reveals two phases of tasks, easy and hard. For easy tasks, our schedule takes the form $\eta(t) = T^{-\xi} f(t/T)$, while for hard tasks our schedule resembles warmup-stable-decay. 
    \item We compare optimal schedules to benchmarks including optimized constant learning rate and optimal powerlaw schedules, finding that our schedules can have better scaling exponents. 
    \item This analysis is generalized to optimal \textbf{ batch size} schedules and optimal \textbf{momentum} schedules. Optimizing batch size and learning rate jointly enables a reduction in total wall-clock time, while optimizing the momentum parameter can provide an improvement in the scaling exponent for hard tasks beyond SGD. 

\item We empirically validate both phases in deep learning settings: ResNets on CIFAR-5M exhibit hard-phase behavior (constant optimal base learning rate under annealing), while GPT-style transformers pretrained on C4 \cite{raffel2020exploring} exhibit easy-phase behavior (optimal learning rate shifts with horizon $T$).
\end{itemize}

\subsection{Related Works}

%Theory scaling laws
Based on the correspondence between neural networks in the lazy training regime \cite{chizat2019lazy}, many works have attempted to characterize neural scaling laws \cite{kaplan2020scaling, hoffmanntraining} by computing the generalization of random feature models with powerlaw features \cite{varre2021last, bordelon2022learning, bahri2024explaining, bordelon2024a, paquette2024, lin2024scaling, ferbach2025dimension}. Several works have examined easy and hard phases of this model in the noise dominated regime \cite{dieuleveut2015non,lin2017optimal}. Notably, \citet{pillaud2018statistical} demonstrated that optimal rates for SGD can be achieved in the hard phase with data repetition and model averaging strategies. \citet{meterez2025seesaw} recently analyzed this model to motivate invariances between learning rate and batch sizes when designing schedules. \citet{qiu2025scaling} identified a scaling collapse phenomenon where the compute optimal models exhibit universal loss dynamics across scales for a variety of learning rate schedules. Our work similarly utilizes this powerlaw random feature model, however we study the \textit{optimal learning rate schedule} for SGD using optimal control theory.

%optimal control approaches to learning

Optimal control theory provides a principled framework for deriving hyperparameter schedules in reduced theoretical models. Pioneering work in the 1990s characterized optimal learning rate schedules for two-layer neural networks within the teacher-student formalism \cite{saad1997globally,rattray1998analysis}. More recently, these techniques have been generalized to several hyperparameter protocols \cite{mori2025optimal,mignacco2025statistical}. Analogous optimization procedures have also recently been studied in deep linear networks \cite{carrasco2023meta} and cognitive science \cite{njaradi2026optimal}.

%theoretical works on learning rate schedules

Most closely related to our work is \cite{li2025unveiling}, which examined learning rate schedules within the context of powerlaw random feature models. However, their analysis was restricted to pre-defined, parametric schedules and did not address the theoretical derivation of the optimal control policy.

\vspace{-5pt}

\section{Solvable Model and Test Loss Dynamics}

We first define the model and describe the evolution of the test error under SGD before formulating the optimal control problem that we aim to solve. 

\subsection{Power Law Linear Random Feature Model}
We consider a random-feature model widely studied in the literature of scaling laws \cite{bordelon2022learning,bordelon2024a,lin2024scaling,paquette2024}.
Let the input data ${\bm x}\in \mathbb{R}^D$ be drawn from a probability distribution $p({\bm x})$. The target function generates noisy labels as
\begin{equation}
    y({\bm x})={\bm w}^*\cdot {\bm \psi}({\bm x})+\sigma_0 z\,,
\end{equation}
where $\sigma_0>0$ is the amplitude of the label noise and $z\sim \mathcal{N}(0,1)$. The vector ${\bm \psi}({\bm x})\in \mathbb{R}^M$ (with $M$ possibly infinite) is a feature map and ${\bm w}^*\in \mathbb{R}^M$ a vector of parameters. We work in the feature eigenbasis, so that the correlation matrix is diagonal
\begin{equation}
  \langle{\bm \psi}_i(\x){\bm \psi}_j(\x)\rangle_{\x\sim p(\x)}=\delta_{ij}\lambda_i\,.
\end{equation}
We consider a linear student model $\hat{y}(\x)={\bm w} \cdot \tilde{\bm \psi}(\x)$, with parameters ${\bm w}\in \mathbb{R}^N$ and projected features $\tilde{\bm \psi}(\x)={\bm G}\bm \psi(\x)$, where ${\bm G}\in \mathbb{R}^{N\times M}$ is a projection matrix. To describe model-size effects, we assume $N\leq M$ and that the top-$N$ features are selected, i.e., ${\bm G}=[{\bm I}_N,{\bm 0}_{N\times (M-N)}]$. We train via online stochastic gradient descent (SGD) with learning rates $\{\eta_t\}$ and batch sizes $\{ m_t \}$ on the square loss
\begin{equation}
\label{eq:SGD}
    \text{SGD Dynamics:} \quad {\bm w}_{t+1} ={\bm w}_t - \frac{\eta_t}{m_t}\sum_{\mu=1}^{m_t} \tilde{\bm \psi}(\x_{\mu,t}) ({\bm w_t}\cdot\tilde{\bm \psi}(\x_{\mu,t}) -y(\x_{\mu,t}))\,.
\end{equation}
The samples $\x_{\mu,t}$ and the label noise variables $z_{\mu,t}$ are sampled independently.

\subsection{Test Loss Dynamics of the Model}
Because this model generates linear dynamics for the parameters, the expected test loss (over random draws of SGD minibatches) can be computed from a simple recurrence. 
\begin{theorem_proper}\label{thm:loss_dynamics}
    Under Gaussian features $\bm\psi$ (or under mild conditions on fourth moments), the expected test loss $L_t$ over the SGD datastream $\mathcal D_{t-1}$ up to time $t$ obeys the following dynamics
    \begin{align}
        L_t = \left< \left( \hat{y}_t(\bm x) - y(\bm x)  \right)^2 \right>_{\x, z, \mathcal D_{t-1}} = \sum_{k=1}^N \lambda_k \underbrace{\left<  (w_{t,k}- w^\star_k)^2  \right>_{\mathcal D_{t-1}}}_{\coloneqq \ c_{t,k}} + \underbrace{\sum_{k=N+1}^M \lambda_k (w^\star_k)^2 +   \sigma_0^2}_{ \coloneqq \ \sigma^2} \label{eq:LT_theorem1}
    \end{align}
    where the mode errors $c_{t,k}$ obey the following \textbf{linear recurrence} that depends on effective noise $\sigma^2$
    \begin{align}
        c_{t+1,k}&=\left(1-2\eta_{t}\lambda_k+\eta_{t}^2 \ \frac{m_t+1}{m_t} \ 
        \lambda_k^2\right)c_{t,k} + \frac{\eta_t^2}{m_t}\lambda_{k}\sum_{\ell=1}^{N} \lambda_{\ell} \  c_{t,\ell}+\frac{\eta^2_t}{m_t}\sigma^2\lambda_k . \label{eq:evolution_ck}
    \end{align}
    This recurrence defines $L_t$ as a function of learning rate and batch size schedules $\{ (\eta_t , m_t) \}$.
\end{theorem_proper}
This result merely extends the main theorem from \cite{bordelon2022learning} to arbitrary learning rate and batch schedules and adds a projection cutoff that generates the effective noise $\sigma^2$. We provide a detailed proof and discussion in Appendix \ref{app:sgd_momentum_derivation}.

% Following \cite{bordelon2022learning}, for $k=1\,\ldots N$, we define $c_{t,k}=\langle (w_{t,k}-w^*_k)^2\rangle_{\mathcal{D}_{t-1}}$, where $\langle \cdot \rangle_{\mathcal{D}_{t-1}}$ indicates the average over all noise sources up to time $t-1$. Assuming that the features ${\bm \psi}({\bm x})$ are zero-mean Gaussian variables, one can show that the coefficients $c_{t,k}$ satisfy the recursion relation \cite{bordelon2022learning}
% \begin{align}
%   c_{t+1,k}&=\left(1-2\eta_{t}\lambda_k+\eta_{t}^2 \frac{m_t+1}{m_t}\lambda_k^2\right)c_{t,k} + \frac{\eta_t^2}{m_t}\lambda_{k}\sum_{\ell=1}^{N} \lambda_{\ell} c_{t,\ell}+\frac{\eta^2_t}{m_t}\sigma^2\lambda_k\,,\label{eq:evolution_ck}
% \end{align}
% where we have defined the total irreducible noise as
% \begin{equation}
% \sigma^2=\sigma_0^2+\sum_{k=N+1}^M(w^{*}_k)^2\lambda_k\,.\label{eq:sigma}
% \end{equation}
% The test loss $L_t$ at time $t$ has the form
% \begin{align}
%  L_t &\coloneqq \langle\langle ({\bm w}^*\cdot {\bm \psi}({\bm x})+ \sigma_0 z -{\bm w} \cdot \tilde{\bm \psi}(\x))^2\rangle_{\mathcal{D}_{t-1}}\rangle_{\x,z} = \sum_{k=1}^N c_{t,k}\lambda_k+\sigma^2\,.
% \end{align}
% It is therefore sufficient to track $c_{t,k}$ in order to predict the test loss dynamics.

\paragraph{Power-law spectra} While the result above holds for arbitrary spectra, we will often specialize to a particular structural assumption on the features. Motivated by empirical observations that natural data often exhibit power-law spectral decay \citep{bordelon2022learning,bahri2024explaining}, we assume power-law scalings for the data covariance eigenvalues and the teacher weights: 
\begin{align}
    \lambda_k \sim k^{-b} \ , \ (w^*_k)^2 \lambda_k \sim k^{-a}
\end{align}
with exponents $a, b > 1$. These exponents $(a,b)$ characterize the effective dimension of the features and difficulty of the learning task. These power-law assumptions correspond to standard source-capacity conditions in the kernel-regression literature, under which the minimax-optimal scaling of the excess risk at fixed batch size is $L_T-\sigma^2\sim T^{-(a-1)/a}$ \cite{caponnetto2007optimal}.

% Under these assumptions, for a constant learning rate $\eta$ and batch size $m$, the excess loss scales as (see Appendix~\ref{app:benchmarks}):
% \begin{equation}
%     L_T - \sigma^2 \sim (\eta T)^{-\frac{a-1}{b}} + \frac{\sigma^2 \eta}{m} \,,
% \end{equation}
% where $\sim$ denotes proportionality up to constant factors. Optimizing over a fixed learning rate yields the optimal constant strategy $\eta \sim T^{-(a-1)/(a+b-1)}$, which results in $L_T - \sigma^2 \sim T^{-(a-1)/(a+b-1)}$. A fundamental question is whether this scaling exponent can be improved by modulating the learning rate over time, and if so, what is the optimal schedule $\eta_T^*(t)$.

\section{Optimal Control for Optimal Hyperparameters} 

From the closed form loss dynamics, we aim to choose the optimal hyperparameter schedules that \textbf{minimize the final loss} $L_T$ after $T$ steps. We can formulate and compare many optimal control strategies with this method. Let $L_t(\bm \eta, \bm m)$ represent the loss for learning rates $\bm\eta = \text{Vec}\{ \eta_t \}_{t=1}^T$ and $\bm m = \text{Vec}\{ m_t \}_{t=1}^T$. Let $\bm 1 \in \mathbb{R}^T$ be the vector of all ones so that $\eta \bm 1$ represents a constant learning rate schedule. Throughout this work, we will define and compare three different optimization problems that correspond to optimizing over different subsets of variables 
\begin{align}
    &\text{Optimal Constant LR:} \quad  \eta^\star_T = \text{argmin}_{\eta \in \mathbb{R}} \quad L_T( \eta \bm 1, m \bm 1 )  \nonumber
    \\
    &\text{Optimal LR Schedule:} \quad  \bm \eta^\star_{T} = \text{argmin}_{\bm\eta \in \mathbb{R}^T }  \ L_T( \bm\eta , m \bm 1 )   \nonumber
    \\
    &\text{Optimal LR and Batch Schedule:} \quad  \bm\eta^\star_T , \bm m^\star_T = \text{argmin}_{\bm\eta \in\mathbb{R}^T ,\bm m \in \mathbb{R}^T }  \ L_T( \bm\eta , \bm m ) .
\end{align}
In the joint LR+batch optimization, the total data budget $B_{\rm tot}=\sum_{t=1}^{T} m_t$ is held fixed. The optimal hyperparameter schedules depend on the total training horizon $T$. We stress that each of these optimization problems is finite dimensional for finite training horizon $T$ and can be computed efficiently due to the linear recurrence for the $c_{t,k}$ variables that define $L_T$ using Theorem \ref{thm:loss_dynamics}. We will take two approaches to these optimization problems. First, we will solve numerically for the optimal learning rate using the \emph{exact recurrence for the loss $L_T$}. Second, we will provide an approximate (yet accurate) description of the optimal control scaling laws under the powerlaw feature assumption.

% \section{Optimal LR schedules via optimal control}

% We formulate the selection of the learning rate schedule as an optimal control problem aimed at minimizing the terminal generalization error $L_T$ for a fixed training budget $T$. Initially, we focus on optimizing the learning rate $\eta_T(t)$ at a constant mini-batch size. Directly applying optimal control to the SGD dynamics in Eq.~\eqref{eq:SGD} is computationally intractable due to the high-dimensional and stochastic nature of the process. Instead, we optimize the control variable acting on the deterministic evolution of the variables $c_{t,k}$ derived in Eq.~\eqref{eq:evolution_ck}. For this analysis, we fix the model size to a large value ($N=1000$ unless stated otherwise). The joint optimization of model size and training time (compute-optimal scaling) is discussed in Section~\ref{sec:compute_optimal}.

\subsection{Numerical optimal control}

\begin{figure*}[htbp]
    \centering
    
    % --- ROW 1: EASY PHASE (b < a) ---
    % Panel (a): Comparison (Easy)

    % Panel (b): Optimal Schedule (Easy)
    \begin{subfigure}[b]{0.32\linewidth}
        \centering
        % Replace with your actual filename for easy phase schedule
        %\includegraphics[width=\linewidth]{figures/schedule_times__N_1000_T_3162_sigma_0.5_a_3.5_b_5.0_batch_False_avgm_5.0_K_200_rho_0.0_etamax_1.0.pdf}
        \includegraphics[width=\linewidth]{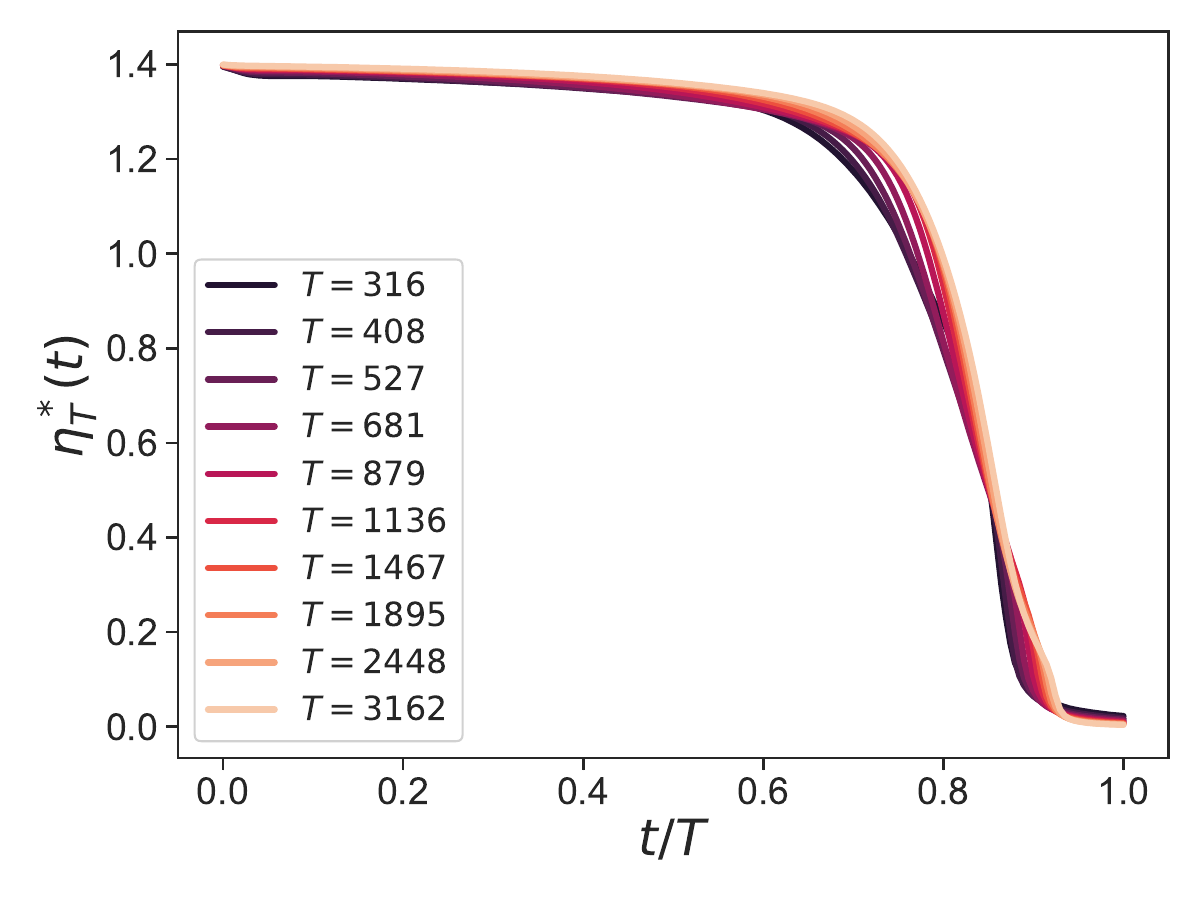}
        \caption{Optimal Schedule ($b > a$)}
    \end{subfigure}
    \hfill
    % Panel (c): Loss vs Time (Easy)
    \begin{subfigure}[b]{0.32\linewidth}
        \centering
        % Replace with your actual filename for easy phase loss
        \includegraphics[width=\linewidth]{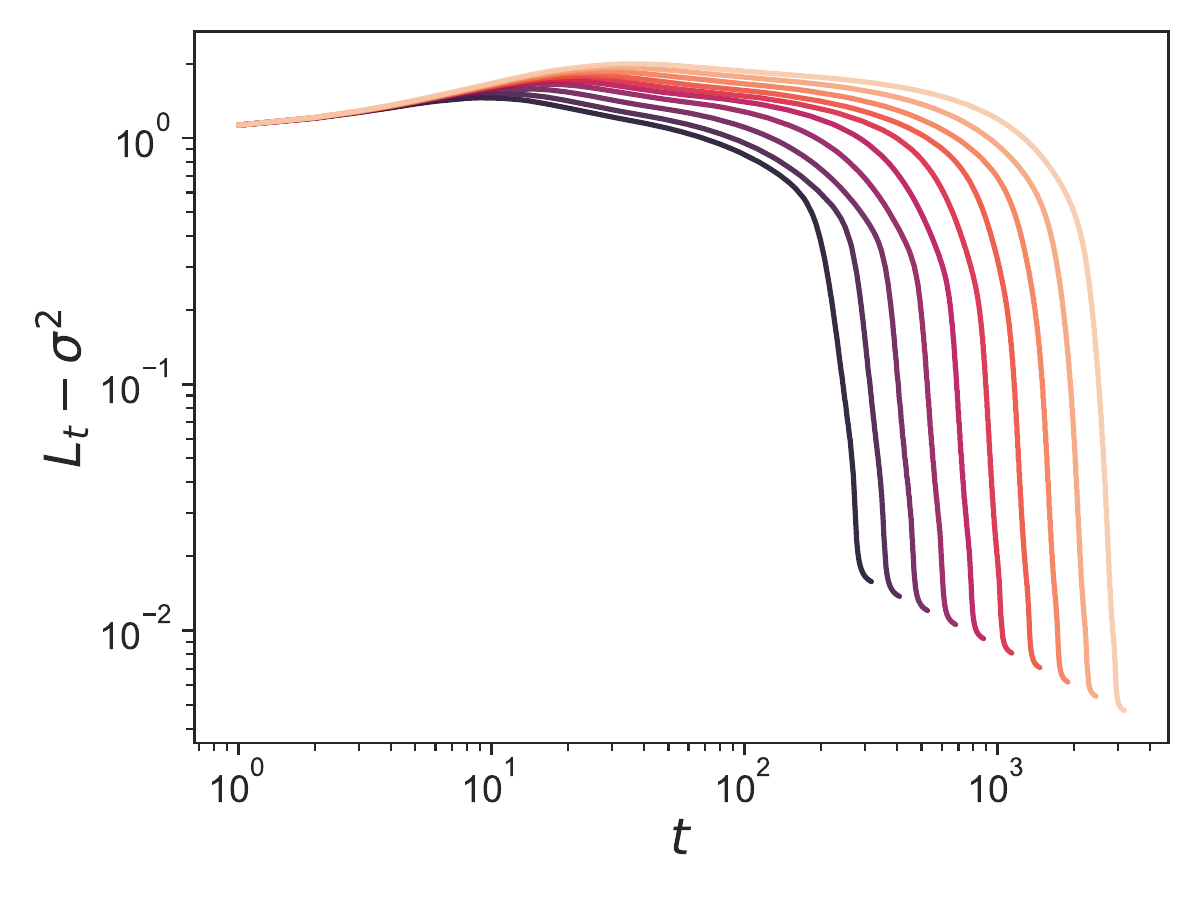}
        \caption{Loss vs. Time ($b > a$)}
    \end{subfigure}
    \hfill
    \begin{subfigure}[b]{0.32\linewidth}
        \centering
        % Replace with your actual filename for easy phase comparison
        \includegraphics[width=\linewidth]{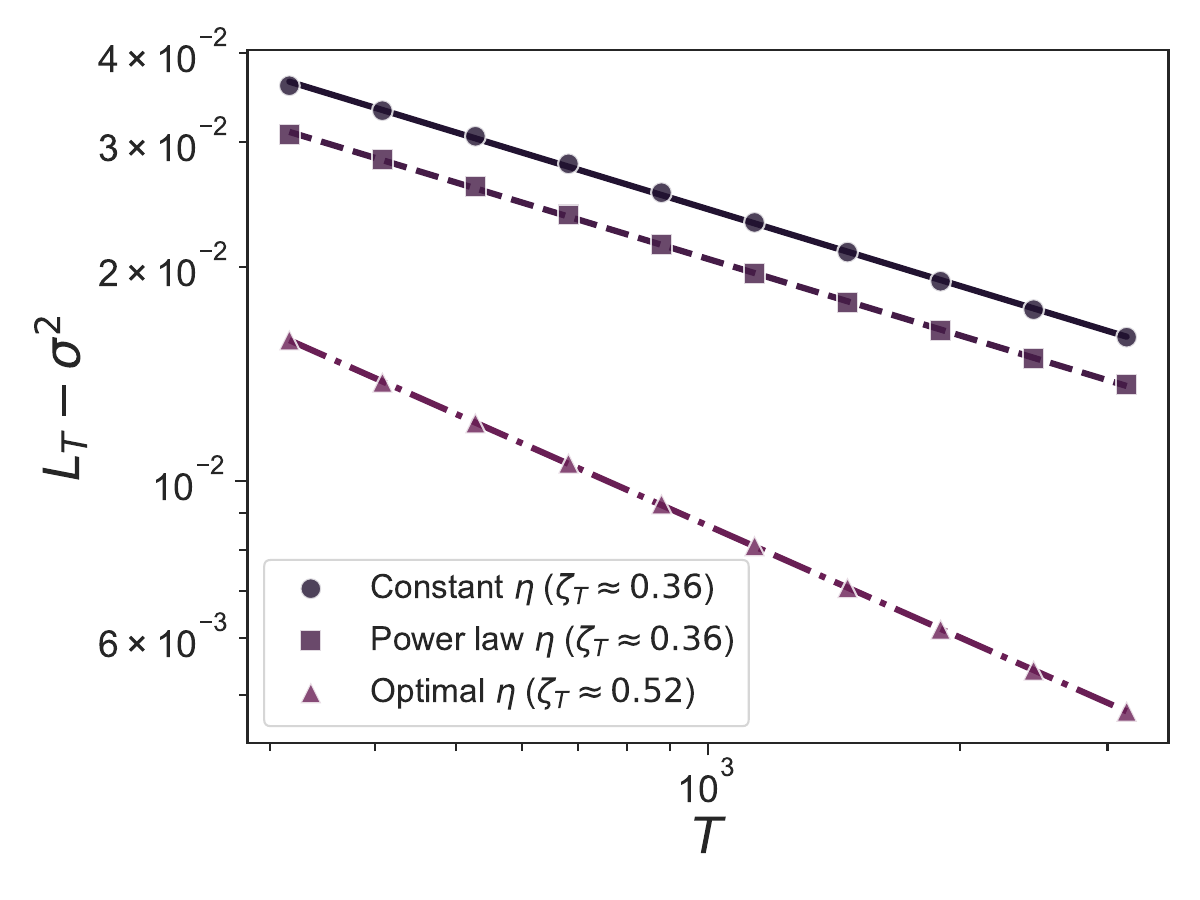}
        \caption{Final Loss ($b > a$)}
    \end{subfigure}
    
    \vspace{1em} % Add vertical space between rows

    % --- ROW 2: HARD PHASE (b > a) ---
    % Panel (d): Comparison (Hard)

    % Panel (e): Optimal Schedule (Hard)
    \begin{subfigure}[b]{0.32\linewidth}
        \centering
        % Replace with your actual filename for hard phase schedule
        \includegraphics[width=\linewidth]{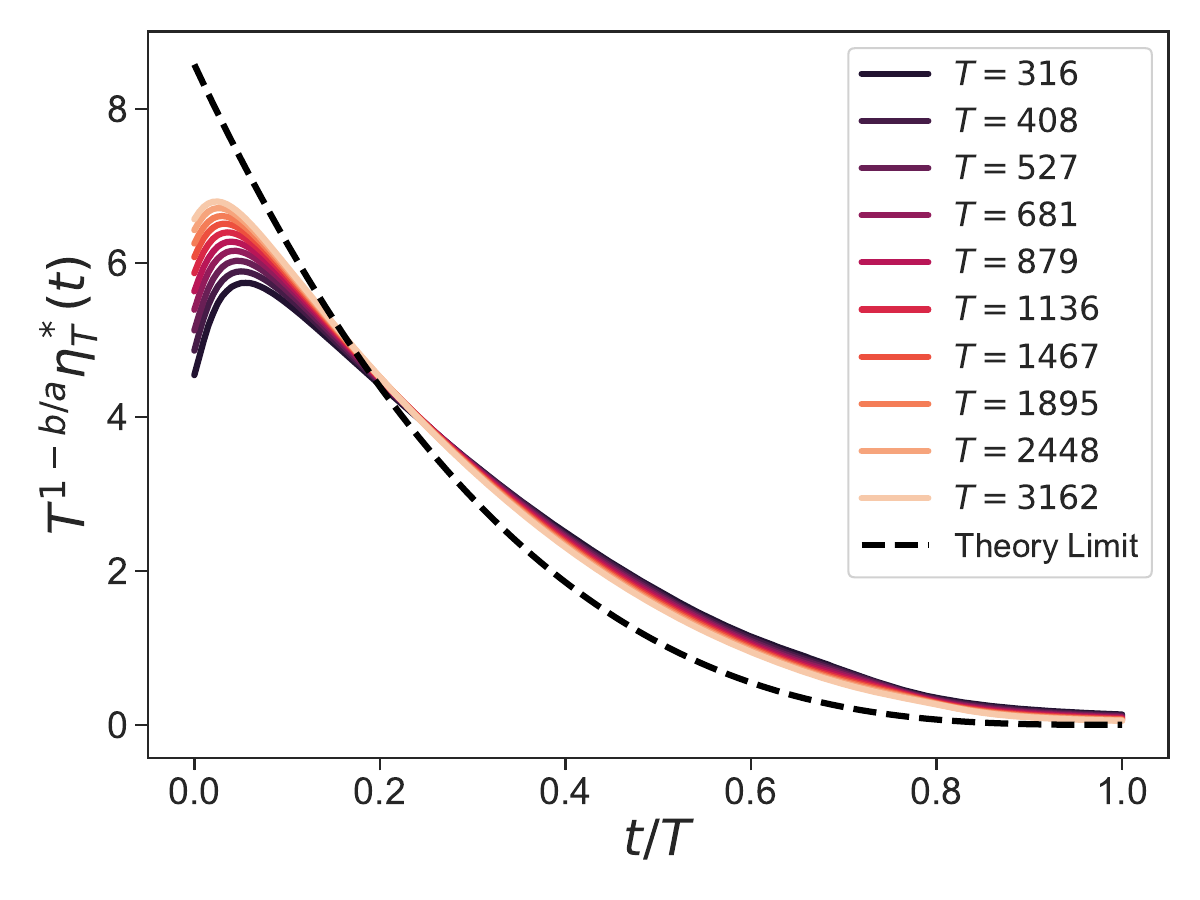}
        \caption{Optimal Schedule ($b < a$)}
    \end{subfigure}
    \hfill
    % Panel (f): Loss vs Time (Hard)
    \begin{subfigure}[b]{0.32\linewidth}
        \centering
        % Replace with your actual filename for hard phase loss
\includegraphics[width=\linewidth]{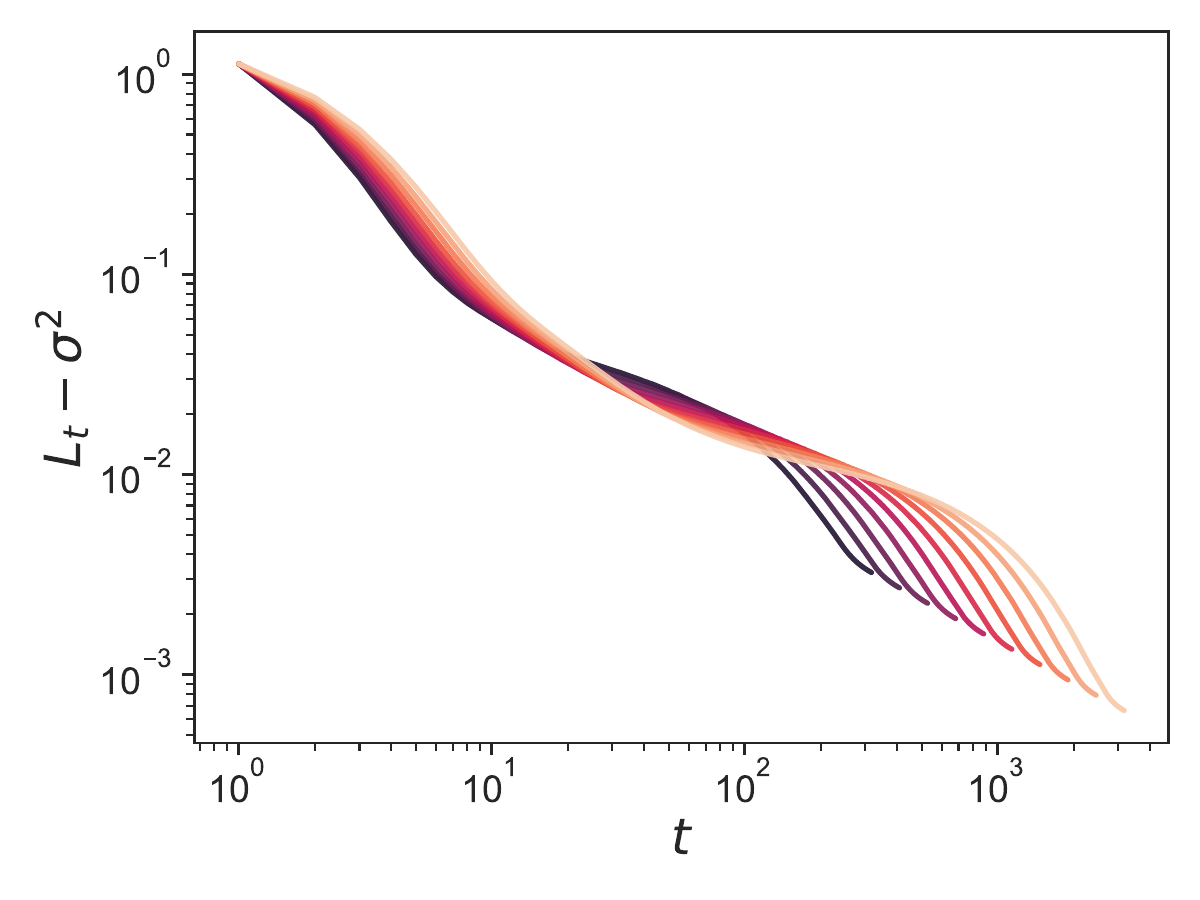}
        \caption{Loss vs. Time ($b < a$)}
    \end{subfigure}
    \hfill
    \begin{subfigure}[b]{0.32\linewidth}
        \centering
        % Replace with your actual filename for hard phase comparison
        \includegraphics[width=\linewidth]{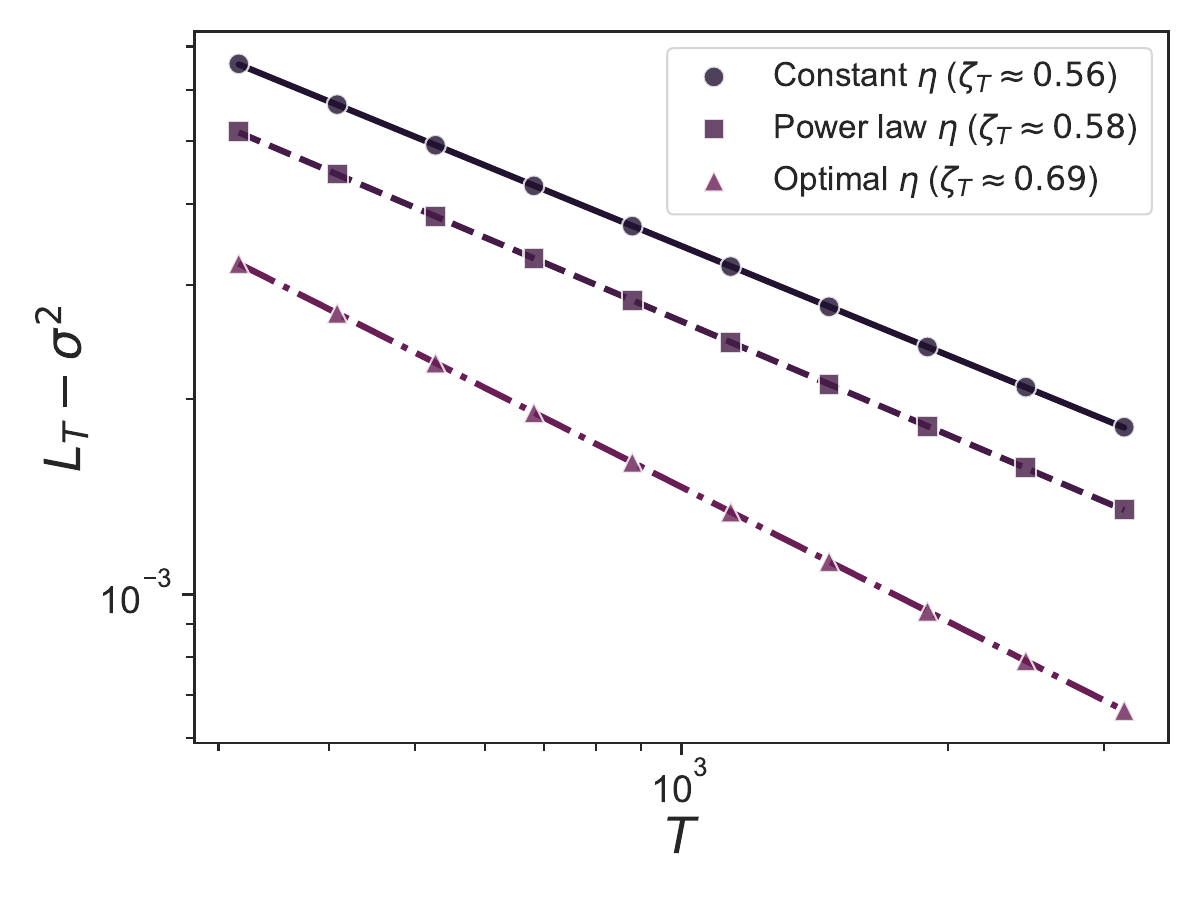}
        \caption{Final Loss ($b < a$)}
    \end{subfigure}
    \caption{\textbf{Comparison of optimal learning rate schedules in the hard ($b > a$) and easy ($b < a$) phases.} 
    \textbf{(a, d)}. Profile of the optimal learning rate $\eta_T^*(t)$. In the hard phase \textbf{(a)}, the schedule maintains an approximately constant maximum value followed by a rapid annealing phase. In the easy phase \textbf{(d)}, the schedule collapses onto the scaling form $\eta_T^*(t) \approx T^{b/a-1}f(t/T)$ (dashed theoretical curve, see Result \ref{thm:opt_LR_powerlaw}). 
    \textbf{(b, e)} Evolution of the loss over training time $t$ for the optimal schedule. \textbf{(c, f)} Scaling of the loss with $T$. The optimal schedule improves scaling exponent compared to the constant and power-law baselines. Exponents obtained from numerical fit match theoretical predictions: for the constant and power-law case, $\zeta_T=\frac{a-1}{a+b-1}$ ($\zeta_T=1/3$ for panel (a) and $\zeta_T=5/9\approx0.56$ for panel (d)), for the optimal schedule $\zeta_T=\min(\frac{a-1}{a},\frac{a-1}{b})$ ($\zeta_T=0.5$ for panel (a) and $\zeta_T=5/7\approx0.71$ for panel (d)).    \textbf{Parameters:} $N=500$, $\sigma=0.5$, $a=3.5$, $m=5$, with $b=5$ (top) and $b=2$ (bottom).  }
    \label{fig:combined_results}
\end{figure*}

We first approach these optimization problems numerically by optimizing the evolution equation in Theorem \ref{thm:loss_dynamics} using a single-shooting method implemented in CasADi \cite{andersson2019casadi}, see Appendix \ref{app:optimal_control} for details. This procedure allows us to identify the optimal learning rate schedules $\eta_T^*(t)$ for various training horizons $T$, as illustrated in Figures~\ref{fig:combined_results}a and \ref{fig:combined_results}d. Crucially, we observe that across different regimes of data structure (parameterized by power-law exponents $a$ and $b$), the optimal schedules achieve a higher scaling exponent compared to the optimal constant-$\eta$ baseline.

\subsection{Analytical derivation of approximate optimal schedules and scaling laws}

To determine the analytical form of the optimal schedule, we consider the continuous-time approximation for large $T$ (when taking this limit, we will often use the notation $\eta(t)$ for a learning rate as a function of time $t \in \mathbb R$). We will provide a comparison of each of the different optimal control problems, starting with optimal constant learning rate (see Appendix \ref{app:benchmarks} for details on the derivation). 

\begin{theorem}[Optimal Constant LR]
    Assuming powerlaw features with exponents $(a,b)$ and nonzero effective noise $\sigma^2 >0$, the late time $T \gg 1$ value of the loss under constant $\eta , m$ is
    \begin{align}
        L_T-\sigma^2 \sim \underbrace{(\eta \ T)^{-\frac{a-1}{b}}}_{\text{Bias}} + \underbrace{\frac{\sigma^2 \eta}{m}}_{\text{Variance}} .
    \end{align}
    As a consequence the optimal learning rate and loss scale as $\eta \sim T^{-\frac{(a-1)}{a-1+b}}$ and $L_T - \sigma^2 \sim T^{-\frac{a-1}{a-1+b}}$. 
\end{theorem} 
This scaling law can be outperformed by optimizing the entire learning rate schedule. 
\begin{theorem}[Optimal LR, Constant Batch]\label{thm:opt_LR_powerlaw}
    Under a continuous time approximation of the loss in Theorem \ref{thm:loss_dynamics} and assuming (1) power-law spectra with exponents $(a,b)$ (2) non-zero effective noise $\sigma^2 > 0$ (3) the batchsize $m(t)=m$ is constant and (4) a maximum learning rate stability condition $\eta(t) \leq \eta_{\max}$
\footnote{The bound $\eta_{\max}=\Theta(\lambda_1^{-1})$ is the leading-mode stability condition $\eta<2/\lambda_1$ of the exact recursion in Theorem~\ref{thm:loss_dynamics}. CasADi enforces it through the dynamics rather than as an explicit constraint, yielding smooth schedules that approximate the analytic plateau-then-decay shape and recover the predicted annealing-fraction scaling (see Fig.~\ref{fig:ts}).}
, the loss for large $T \gg 1$ under the optimal learning rate schedule minimizes the following functional of the integrated learning rate $\chi(t)$
    \begin{align}
        L_T -\sigma^2 \sim \underbrace{\chi(0)^{-\frac{a-1}{b}}}_{\text{Bias}} + \underbrace{\frac{\sigma^2 }{m} \int_0^T dt \ \dot\chi(t)^2 \ 
        \chi(t)^{-2 + \frac{1}{b}}}_{\text{Variance}}
        \quad  \quad  \underbrace{\chi(t) \coloneqq \int_t^T ds \  \eta(s) }_{\text{Integrated Learning Rate}}
    \end{align}
    From Noether's theorem \cite{gelfand2000calculus}, the optimal schedule satisfies $\dot\chi(t) \propto - \chi(t)^{(2b-1)/2b}$ with boundary condition $\chi(T) = 0$. The optimal solution exhibits two phases. In the \textbf{Easy Phase} ($a > b$), the optimal learning rate schedule is 
    \begin{align}
       \text{Easy Phase:} \quad \eta_T^*(t) \sim  c(a,b) \ \   T^{b/a-1} \  f\left( \frac{t}{T} \right) \ , \ f(z) = (1-z)^{2b-1}
    \end{align}
    where $c(a,b)$ is a constant. In the \textbf{Hard Phase} ($a<b$), the optimal learning rate saturates the stability bound $\eta = \eta_{\max}$ for early times $t < t_s$ before subsequently annealing
    \begin{align}
        \text{Hard Phase:} \quad \eta_T^*(t) \sim \begin{cases}
        \eta_{\max} & t < t_s\,,\\
        \eta_{\max}\left(\frac{1-t/T}{1-t_s/T}\right)^{2b-1} & t > t_s\,,
    \end{cases}
    \end{align}
    where the fraction of time spent annealing scales sublinearly with $T$ as $1-t_s/T \sim T^{-\frac{b-a}{2b-1}}$. For derivations and exact constants, consult Appendix \ref{app:long_time_dynamics}. 
\end{theorem}

Our analysis revealed two distinct scaling regimes depending on the parameters $a$ and $b$. Following the terminology of \cite{pillaud2018statistical}, we distinguish between an \textit{easy phase} ($b < a$), where the constraint is asymptotically inactive, and a \textit{hard phase} ($b > a$), where it plays a critical role\footnote{We note that this easy/hard distinction is focused on the behavior of label noise SGD and differs from other notions of easy and hard phases such as diverging RKHS norm \cite{paquette2024, bordelon2025feature}.}. The type of schedule that we identify as optimal in the easy phase is commonly known as \emph{polynomial decay} and has been used in a variety of applications \cite{wu2023selecting}, while the optimal protocol in the hard phase closely resembles a \emph{warmup-stable-decay} (WSD) schedule \cite{hu2024minicpm} \footnote{The optimal hard-phase schedule consists of a constant phase followed by a decay phase and has no warmup. We retain the WSD label because it is the more standard term in the literature for schedules of this form.}
. We next show that the scaling exponents for optimized schedules are improved relative to the constant-$\eta$ baseline. In Figure \ref{fig:combined_results} we illustrate these two phases and plot both the exact solution to the optimal control problem, obtained numerically, and show that our scaling exponents derived from large $T$ asymptotics accurately capture the scaling laws for each schedule. 

\begin{corollary}[Scaling Law Under the optimal LR Schedule]
    The scaling law of the loss with $T$ is 
    \begin{align}
        L_T - \sigma^2 \sim T^{- (a-1)/\max(a,b)}\label{eq:scaling_laws_optimal}
    \end{align}
which outperform the scaling exponents for optimal constant LR. In the easy phase, this rate matches the minimax-optimal exponent $T^{-(a-1)/a}$ \cite{caponnetto2007optimal}.
\end{corollary}

\paragraph{Interpretation} In the easy phase ($a>b$), the optimal schedule balances the contribution of the bias and variance components. In the hard phase ($b > a$), the slow decay of the bias implies that the unconstrained optimal learning rate would grow as $T^{b/a-1}$ for large $T$. This divergence activates the stability constraint $\eta_T(t) \le \eta_{\max}$, resulting in the WSD profile. As shown in Figure \ref{fig:loss_decomposition}, while the easy phase schedule gradually decays to minimize both error components simultaneously, the hard phase strategy dedicates the majority of the training budget to maximizing bias suppression via a constant, maximal learning rate, compressing the variance-reducing annealing into a vanishing fraction of the total training time. In both regimes, a critical implication is that the optimal schedule relies on a fixed training horizon $T$. Anytime protocols, which are agnostic to the total budget, cannot effectively postpone variance minimization. We summarize all of the exponents for these settings in Table \ref{tab:benchmarks}.  The advantage of horizon-dependent schedules has been previously observed in convex optimization \cite{jain2019making,defazio2023optimal}. Previous works have rationalized the effectiveness of the WSD schedule by conjecturing a ``river valley'' loss landscape \cite{wen2024understanding,liu2025neural}. Here, we demonstrate that WSD is the optimal strategy in the hard phase, providing an alternative explanation for its effectiveness based on balancing demands of bias and variance reduction. The scaling laws in Eq.~\eqref{eq:scaling_laws_optimal} can be combined with the dependence of $\sigma^2$ on the model size $N$ to derive the compute-optimal allocation of $N$ and $T$ at fixed compute $C=mNT$; see Appendix~\ref{app:compute_optimal}.

\subsection{Comparison with benchmarks}

\begin{table}[h]
    \centering
    \begin{tabular}{|c|c|c|c|c|c|}
            \hline
        Phase &  Constant LR & Powerlaw & Opt. LR Schedule & Opt. LR + Batch Schedule \\
        \hline
        Easy $(a>b)$ &  $\frac{a-1}{a-1+b}$ & $\frac{a-1}{a-1+b}$ & $\frac{a-1}{a}$ & $\frac{a-1}{b}$ \\ 
        \hline
        Hard $(a<b)$ & $\frac{a-1}{a-1+b}$ & $\frac{a-1}{a-1+b}$ & $\frac{a-1}{b}$ & $\frac{a-1}{b}$ \\
        \hline
    \end{tabular}
    \caption{Comparison of scaling exponents $\zeta_T$ for the loss as a function of training horizon $L_T - \sigma^2 \sim T^{-\zeta_T}$ for different types of schedules. Constant LR refers to $\eta_T(t) \sim T^{-\xi}$ and powerlaw refers to $\eta_T(t) \sim T^{-\xi} t^{-\delta}$, where the exponents $\xi$ and $\delta$ are chosen optimally in each case, and optimal is the schedule we derive from optimal control. In the easy phase the batch schedule improves the wall-clock exponent from $\frac{a-1}{a}$ to $\frac{a-1}{b}$, while leaving the sample-budget exponent $\frac{a-1}{a}$ unchanged.}
    \label{tab:benchmarks}
\end{table}

Having established that the optimal schedule improves the scaling exponent relative to any constant learning rate strategy, we extend our comparison to a broader class of protocols. In Appendix~\ref{app:benchmarks}, we analyze power-law schedules of the form $\eta_T(t)\sim T^{-\xi}t^{-\delta}$, optimizing the exponents $\xi$ and $\delta$ to minimize the final loss. We demonstrate that even the best-performing power-law schedule asymptotically recovers the same scaling behavior as the optimal constant-$\eta$ strategy. Consequently, our derived optimal schedule strictly outperforms the entire family of power-law benchmarks (see Table \ref{tab:benchmarks}). Our schedules saturate information theoretically optimal rates in the easy phase, but not the hard phase where $b > a$ \cite{fischer2020sobolev}. \citet{pillaud2018statistical} showed that a combination of data repetition and model averaging can improve the rates in the hard phase to saturate the theoretically optimal rate.

Generalizing beyond the specific optimal schedule derived in Result \ref{thm:opt_LR_powerlaw}, we investigate the sensitivity of the performance to the choice of the profile function within the general scaling $\eta_T(t)= T^{-\xi} g\left(t/T\right)$. In Appendix~\ref{app:benchmarks}, we show that the specific shape of $g(z)$ does not change the scaling exponent, provided it decays faster than $(1-z)^{b-1}$ as $z\to 1$. The prefactor exponent $\xi$ must be chosen as $\xi=\max[0,1-b/a]$ (using a different value would result in suboptimal scaling). Under these conditions, the generic schedule recovers the optimal scaling exponents in both the easy and hard regimes, differing from the optimal error only by a constant prefactor.

\subsection{Joint optimization of learning rate and batch size}
Our analysis extends naturally to the joint optimization of the learning rate $\eta(t)$ and minibatch size $m(t)$ under a fixed data budget $B_{\rm tot}\!=\!\int_0^T dt~ m(t)$, with the wall-clock time $T$ as a free variable.

\begin{theorem}[Joint LR and Batch Schedules]\label{thm:joint_LR_batch}
Under the assumptions of Result~\ref{thm:opt_LR_powerlaw}, optimizing $L_T$ jointly over $\eta(t)$ and $m(t)$ subject to $B_{\rm tot}=\int_0^T dt\,m(t)$ yields a degenerate family of optima with $m(t)\propto -\dot\chi(t)\,\chi(t)^{-1+1/(2b)}$, all reaching the same loss
\begin{align}
    L_T - \sigma^2 \sim B_{\rm tot}^{-(a-1)/a}\;\;(b<a)\,, \qquad L_T - \sigma^2 \sim B_{\rm tot}^{-(a-1)/b}\;\;(b>a)\,.
\end{align}
The degeneracy is broken, following \citet{meterez2025seesaw}, by demanding minimum wall-clock time, which fixes $\eta(t)=\eta_{\max}$ and yields an increasing batch schedule
\begin{align}
    m^*_T(t) = \frac{B_{\rm tot}}{2bT}\left(1-\frac{t}{T}\right)^{1/(2b)-1}\,,\qquad T \sim B_{\rm tot}^{b/a}\;\;(b<a)\,,\quad T \sim B_{\rm tot}\;\;(b>a)\,.
\end{align}
\end{theorem}

This type of procedure, known as a \emph{batch ramp}, has been empirically proposed \cite{smith2017don} but is derived here through optimal control principles. In the easy phase, the batch ramp accelerates training in wall-clock time: the loss decays as $L_T-\sigma^2\sim T^{-(a-1)/b}$ rather than $T^{-(a-1)/a}$ at constant batch, while the sample-budget exponent $(a-1)/a$ is unchanged (see Fig.~\ref{fig:batch_ramp}). In the hard phase, both the wall-clock and sample-budget exponents coincide with the fixed-batch optimum.

%Compare regimes where there is a irreducible noise floor to cases where only noise comes from model size ($\sigma^2_0 = 0$)? Multiplicative noise becomes more important

%\textcolor{red}{does our schedule achieve supercollapse?}

\subsection{Optimal Momentum Schedules Compared to SGD}

\begin{figure*}
    \centering
    \begin{subfigure}[b]{0.32\linewidth}
        \centering
    \includegraphics[width=\linewidth]{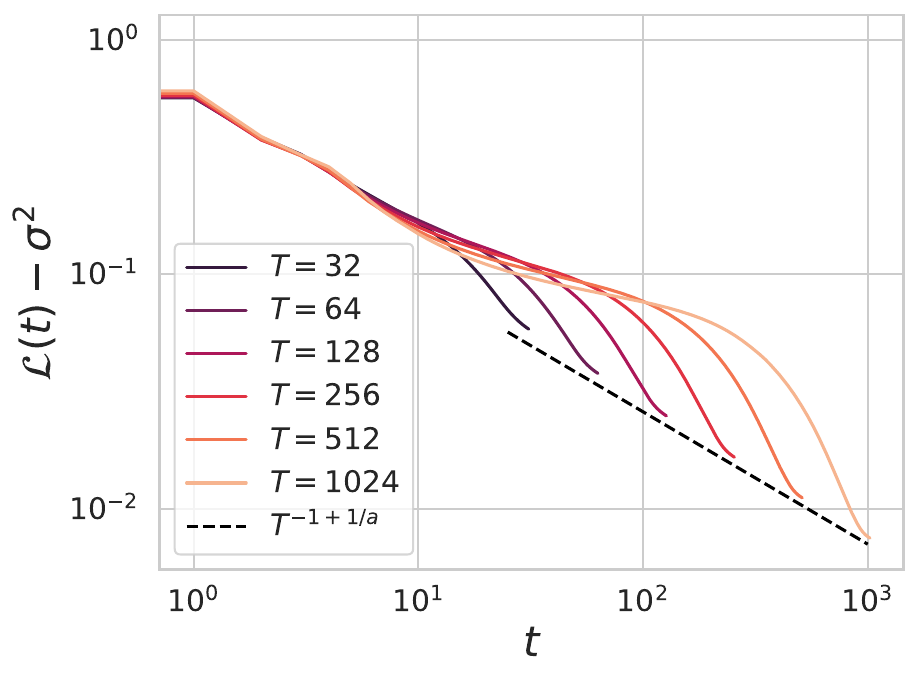}
    \caption{Easy-Phase Loss Dynamics}
    \end{subfigure}
    \begin{subfigure}[b]{0.32\linewidth}
        \centering
    \includegraphics[width=\linewidth]{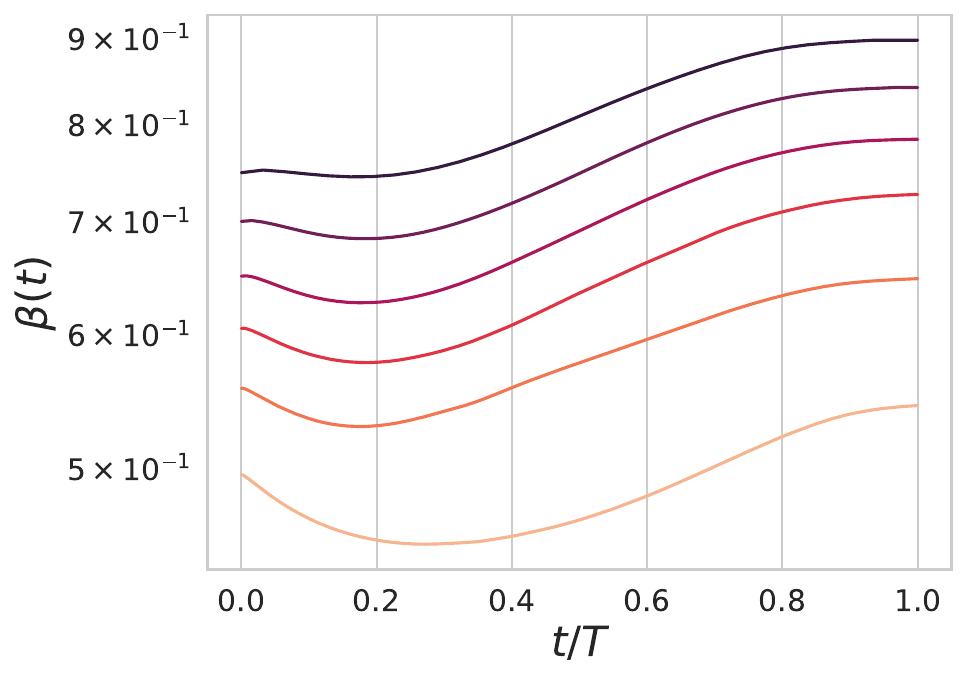}
    \caption{Easy-Phase Momentum Schedule}
    \end{subfigure}
    \begin{subfigure}[b]{0.32\linewidth}
        \centering
    \includegraphics[width=\linewidth]{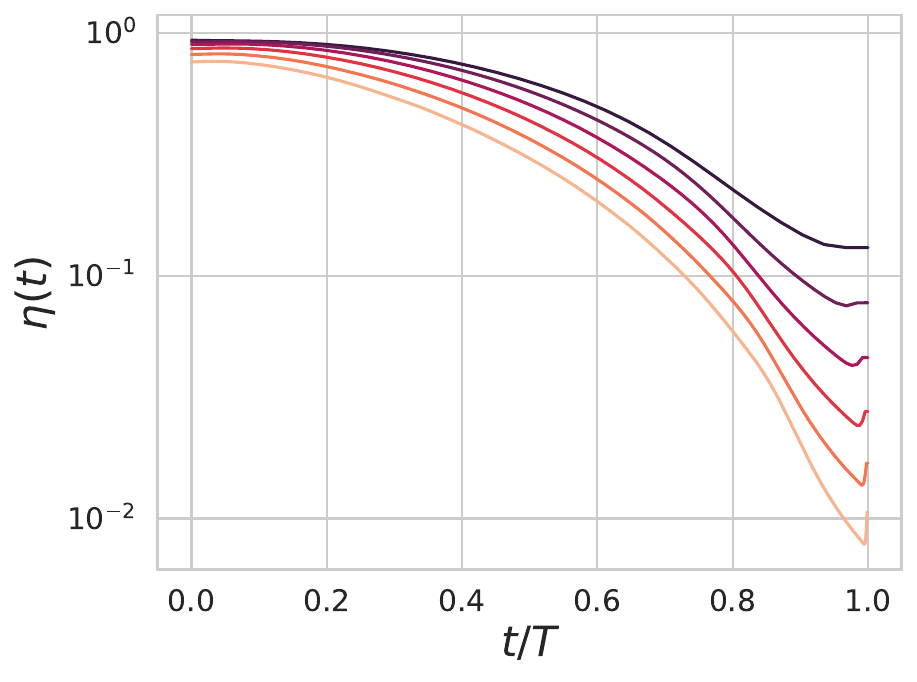}
    \caption{Easy-Phase Learning Rate}
    \end{subfigure}
   \begin{subfigure}[b]{0.32\linewidth}
        \centering
        \includegraphics[width=\linewidth]{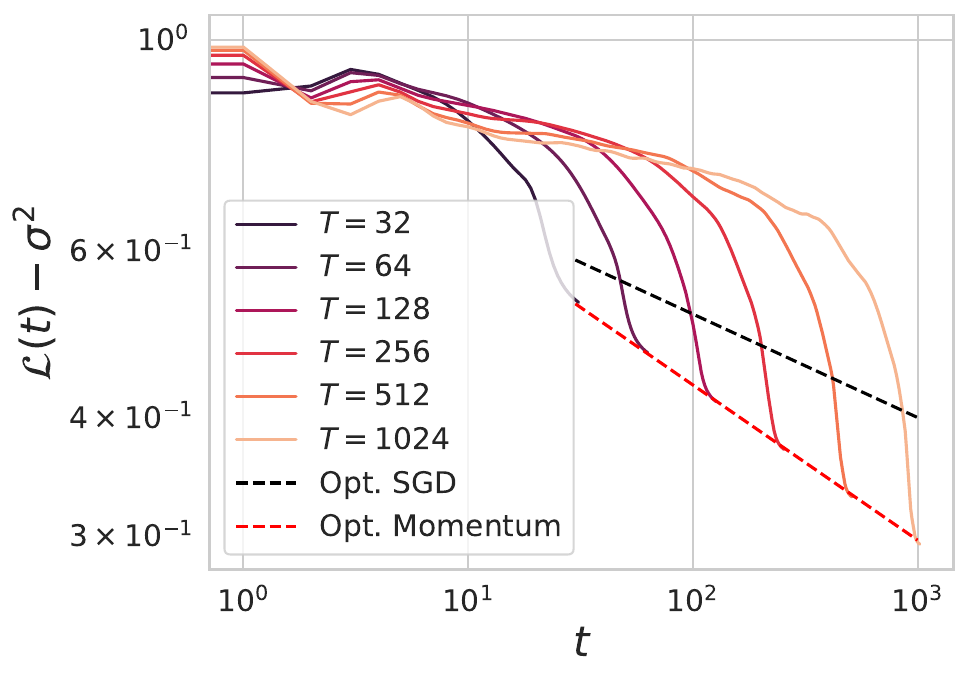}
        \caption{Hard-Phase Loss Dynamics}
    \end{subfigure}
    \begin{subfigure}[b]{0.32\linewidth}
        \centering
        \includegraphics[width=\linewidth]{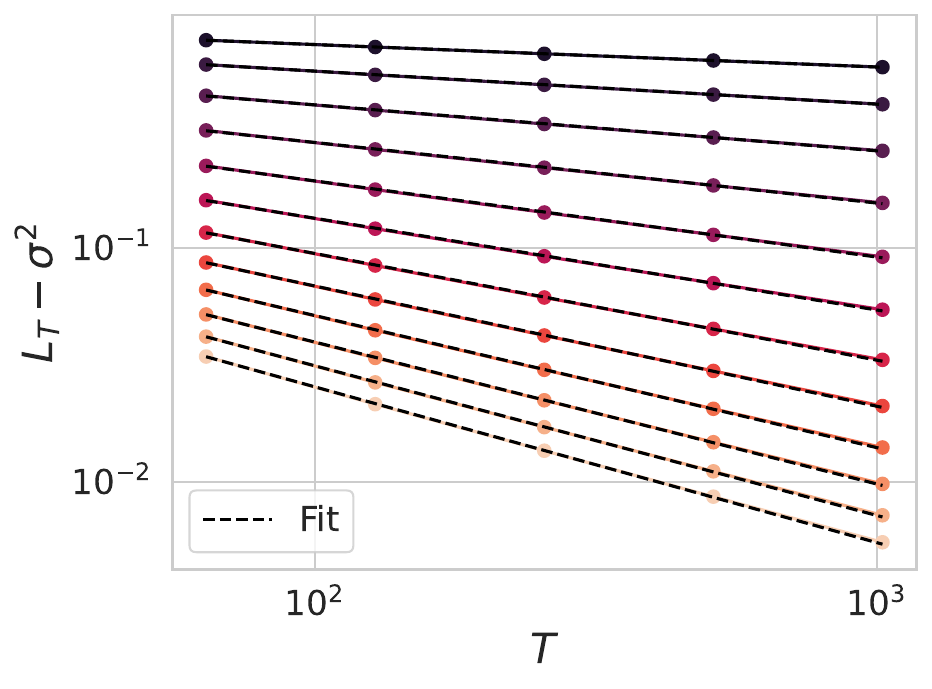}
        \caption{Opt. Momentum Varying $a/b$}
    \end{subfigure}
    \begin{subfigure}[b]{0.32\linewidth}
        \centering
        \includegraphics[width=\linewidth]{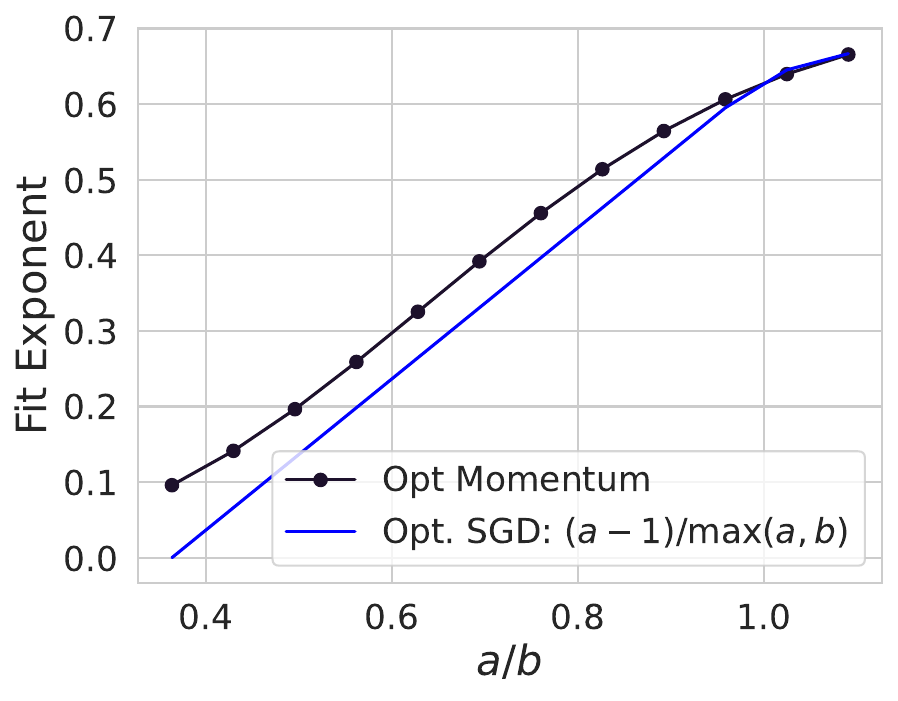}
        \caption{Improved Scaling in Hard Phase}
    \end{subfigure}
    \caption{Optimal Schedule and loss dynamics for SGD + momentum. (a) For the easy task regime $(a,b)=(2.3,1.5)$, the numerically optimized schedule achieves the same scaling law as SGD with optimal schedule $L_T - \sigma^2 \sim T^{-1+1/a}$. (b) The optimal momentum dynamics vary significantly across $T$ but only weakly vary with $t$. (c) The learning rate for optimal momentum schedules anneals similarly to SGD in the easy phase. (d) In the hard phase $(a,b)=(1.3,2.8)$, the scaling law for the loss obtained by jointly optimizing momentum and learning rate is better than the SGD rate $T^{-(a-1)/b}$. (e)-(f) We fit scaling laws for the optimal momentum schedule $L_T -\sigma^2 \sim T^{-\xi}$ for varying $(a,b)$ and plot the fit exponent compared to our predictions for SGD. For $a<b$, momentum outperforms SGD.}
    \label{fig:opt_momentum_schedules}
\end{figure*}

\begin{figure*}
    \centering
    \begin{subfigure}[b]{0.32\linewidth}
        \centering\includegraphics[width=\linewidth]{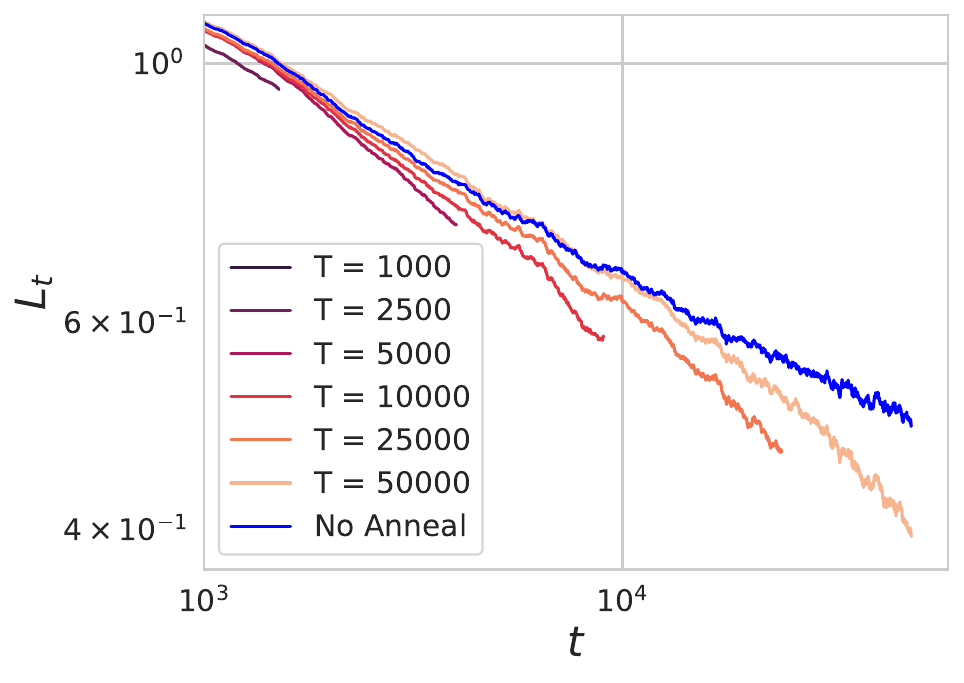}
        \caption{Annealing vs Fixed LR }
    \end{subfigure}
    \begin{subfigure}[b]{0.32\linewidth}
         \centering\includegraphics[width=\linewidth]{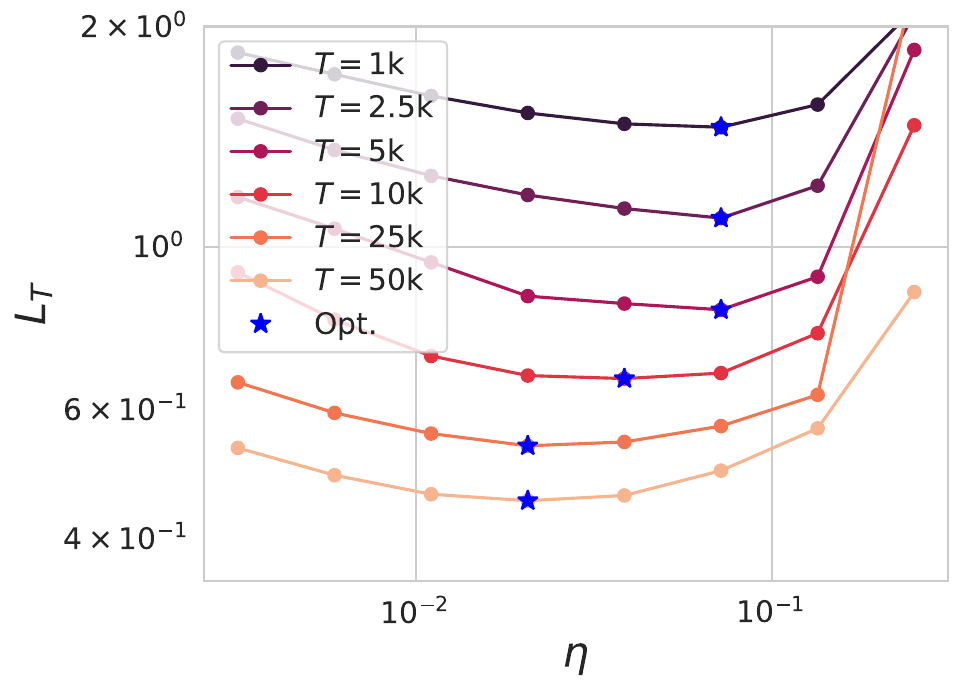}
         \caption{Fixed LR Doesn't Transfer}
    \end{subfigure}
    \begin{subfigure}[b]{0.32\linewidth}
        \centering\includegraphics[width=\linewidth]{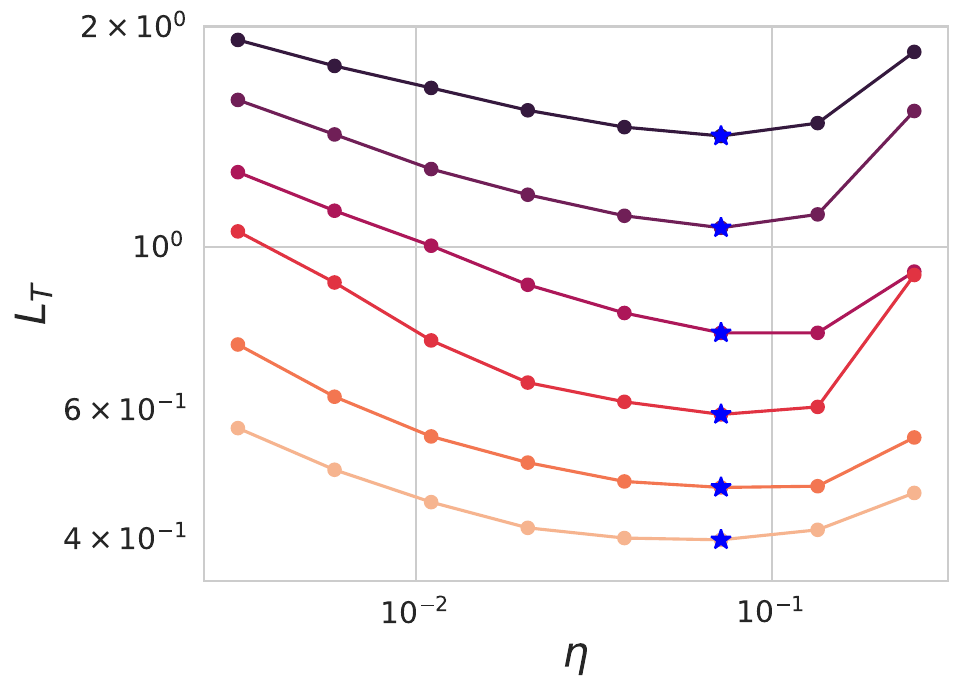}
        \caption{ LR Transfer with Annealing }\end{subfigure}

    \begin{subfigure}[b]{0.32\linewidth}
        \centering\includegraphics[width=\linewidth]{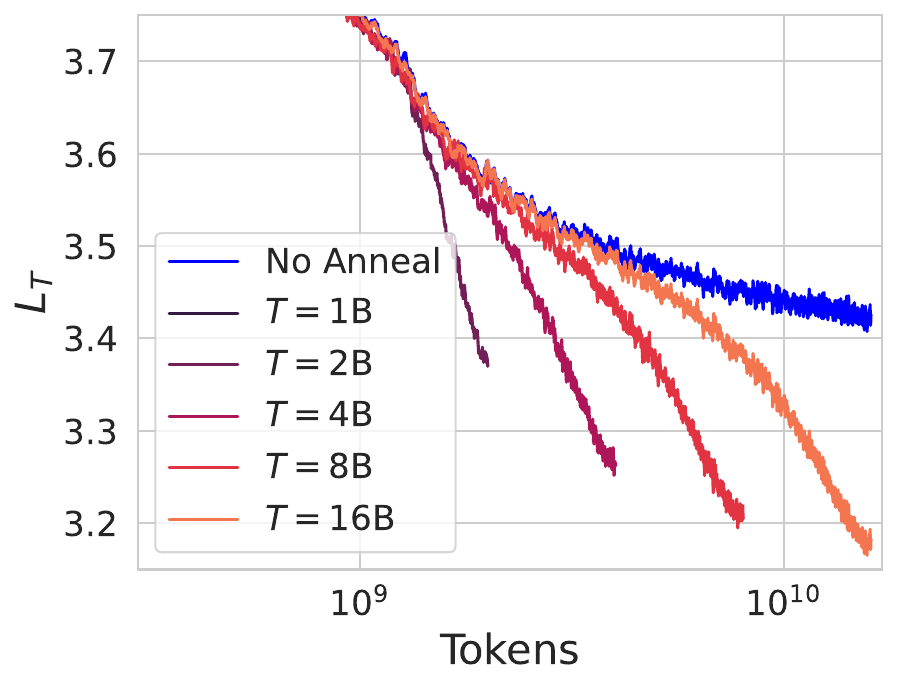}
        \caption{LLM Constant vs Annealing}
    \end{subfigure}
    \begin{subfigure}[b]{0.32\linewidth}
        \centering\includegraphics[width=\linewidth]{figures/const_opt_LR_shift_OLMO_125M.pdf}
        \caption{Fixed LR}
    \end{subfigure}
    \begin{subfigure}[b]{0.32\linewidth}
        \centering\includegraphics[width=\linewidth]{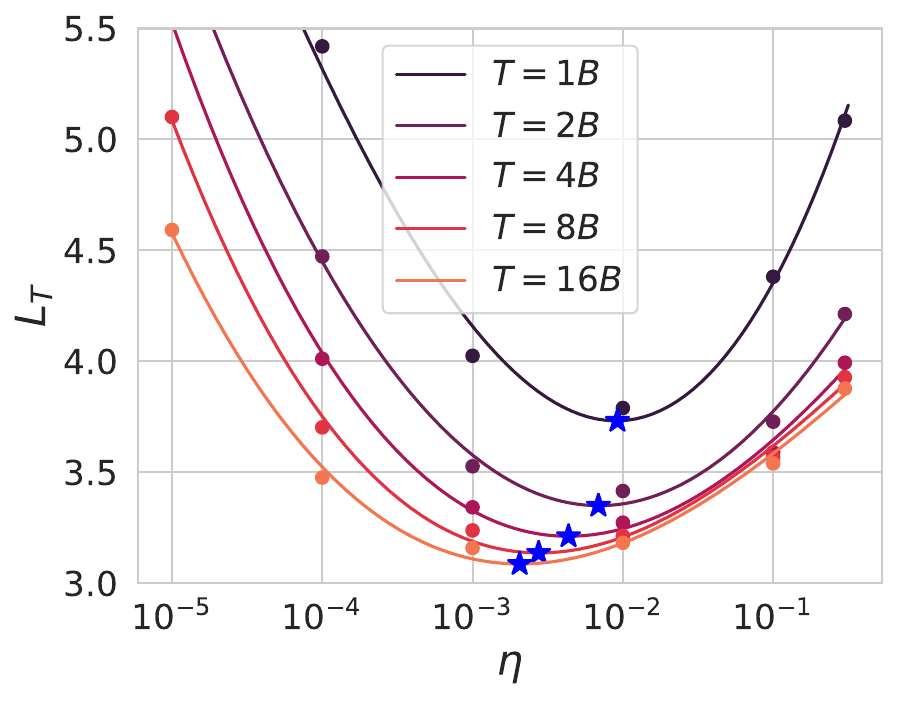}
        \caption{Annealing Needs Tuned Base LR}
    \end{subfigure}
    \caption{Hard and easy phase behaviors in different deep learning settings. (a) ResNet CNNs trained with SGD on CIFAR-5M. Cross-entropy loss dynamics as a function of training time $t$ for fixed learning rate (blue) and polynomial annealing schedule set to training horizon $T$. The final loss follows a better trend when using the annealing schedule. (b) Loss as a function of $(\eta, T)$ and for fixed learning rate. The optimal learning rate decreases with $T$. (c) For polynomial annealing schedules, the optimal base learning rate is approximately constant across horizons $T$, consistent with our model in the hard phase. Further, the best performance achieved at $T$ iterations is noticeably improved compared to constant learning rate. (d)-(f) The same plots but for GPT-2 transformer pretraining on C4. In this case, both the fixed learning rate and annealing strategies require decreasing the base learning rate with $T$.  }
    \label{fig:cifar5m_lr_transfer}
\end{figure*}
We also study momentum with time varying learning rates $\{ \eta_t \}$ and momentum $\{ \beta_t \}$ parameters which generate the update equations for parameters $\bm w_t$ 
\begin{align}
    &\w_{t+1} = \w_t + \eta_t \bm v_t \nonumber \  , \ \bm v_t = (1-\beta_t) \bm v_{t-1} + \beta_t \bm g_t \nonumber
    \ , \ \bm g_t \coloneqq \frac{1}{m_t} \sum_{\mu=1}^{m_t} \tilde{\bm\psi}(\x_{\mu,t}) \left( y(\x_{\mu,t}) - \hat{y}(\x_{\mu,t}) \right)\,.
\end{align}
We optimize the terminal loss $L_T$ over the choices of $\{ (\eta_t, \beta_t ) \}_{t=1}^T$. Using similar methods to the analysis we used for SGD in Theorem \ref{thm:loss_dynamics}, we can compute the test loss by tracking three correlation variables $\{ C_{k,t}, R_{k,t}, V_{k,t} \}$ for each eigendirection $k$. These variables are defined as
\begin{align}
    C_{k,t} = \left< (w_{k,t}-w^\star_k)^2 \right> \,, \quad \quad \quad   
    R_{k,t} = \left< v_{k,t} (w^\star_k - w_{k,t})  \right>  \,, \quad \quad \quad  V_{k,t} = \left< (v_{k,t})^2 \right> 
\end{align}
 which collectively evolve as a linear dynamical system that depends on $\lambda_k, w^\star_k,\eta_t, \beta_t$ and $m$. From this dynamical system, we can obtain the loss dynamics $L_T - \sigma^2 = \sum_k \lambda_k C_{k,T}$, enabling use of optimal control. Recent work has demonstrated that allowing $\beta$ to depend on time can improve scaling laws in this powerlaw random feature model \cite{ferbach2025dimension}. Instead of a preset schedule for $\beta_t$, we investigate the \textit{optimal control} policy for $\eta_t$ and $\beta_t$. We plot the result for easy and hard tasks in Figure \ref{fig:opt_momentum_schedules}, finding it differs from the anytime powerlaw schedule used by \cite{ferbach2025dimension}. We find that the optimal control learning rate for easy tasks exhibits phenomenology remarkably similar to optimal SGD. However, for hard tasks, we find that optimizing the momentum schedule $\beta_t$ in addition to the learning rate schedule $\eta_t$ provides an improvement to the scaling law with $T$ in the hard phase ($a<b$). The optimal momentum variables $\beta_t$ in either case show an initial transient decrease before a final increase as $t/T \to 1$.

\section{Experiments in Deep Networks}

\paragraph{Computer Vision} We study a simple setup of CIFAR-5M (an extended version of CIFAR-10 that allows one-pass training) image classification in a deep convolutional ResNet \cite{nakkiran2020deep}. We train with SGD using either fixed learning rate or a polynomial decay schedule, consistent with our optimal schedule. In Figure \ref{fig:cifar5m_lr_transfer}, we plot a comparison of the learning rate transfer for fixed learning rate schedules compared to polynomial decay. In Figure \ref{fig:cifar5m_lr_transfer} (a) we show that the loss dynamics $L_t$ for fixed learning rate (blue) can be outperformed at the same base learning rate by annealing with a schedule that decays to zero as $t/T \to 1$. In Figure \ref{fig:cifar5m_lr_transfer} (b), we plot the optimal learning rate as a function of $T$ for constant learning rate, which shifts left. Lastly, when using an annealing schedule we find that the optimal learning rate is approximately constant and the best performance at each value of $T$ is improved compared to fixed learning rate. These observations are consistent with the behavior of our theoretical model in the hard phase since the optimal base learning rate of an annealing schedule is approximately constant.  Based on the spectral analysis of the NTK for ResNets on CIFAR-5M from \cite{bordelon2024a}, it is reasonable to expect this problem-architecture pair to fall in the hard phase. We note that our theory \textit{does not generally predict transfer of base learning rates}, since in the easy phase, the optimal base learning rate will still scale as a powerlaw in $T$.

\paragraph{Language Modeling} We also find evidence for easy-phase behavior in GPT-style transformers trained on next word prediction on natural language. The models have 125M non-embedding parameters and are trained on the C4 corpus \cite{raffel2020exploring} with Adam with varying learning rate and default $\beta$. In Figure \ref{fig:cifar5m_lr_transfer} (d) we plot the influence of cosine annealing schedules on the loss dynamics at the near optimal learning rate. In Figure \ref{fig:cifar5m_lr_transfer} (e)-(f) we illustrate that for both fixed learning rate and annealing, the optimal base learning rate decreases with $T$. The persistence of a shift in the optimal base learning rate under annealing is compatible the \emph{easy phase} in our theory.

\section{Conclusions}

We theoretically analyzed the optimal learning rate schedule for a simple random feature model. Our results utilized optimal control methods, which enabled optimization over the full trajectory of the learning dynamics. This model recapitulated schedules often observed in practice including polynomial decay schedules and warmup-stable-decay. Further, we explored optimal batch size and optimal momentum schedules, finding improvements in wall-clock scaling and sample-budget scaling, respectively.

\vspace{-10pt}
\paragraph{Limitations and Future Directions}
While our analysis provided some insights into the behavior of SGD under different annealing schedules and different phases of task difficulty, there are several limitations to the present work. First, our current theory is focused on describing linear models trained with SGD.  This can only capture optimization behaviors of neural networks operating in the lazy/kernel regime and also fails to capture the behavior of more sophisticated optimizers such as Adam or preconditioned optimizers like Muon, Shampoo, or SOAP \cite{diederik2014adam, jordan2024muon, gupta2018shampoo, vyas2024soap}. Second, while our numerical solution to optimal momentum revealed an improvement in the scaling law exponent, we have not yet identified analytically what this improved exponent is for the hard phase. Lastly, while we provided experimental evidence for easy and hard phase problems in different deep learning settings, it would be interesting to develop ways to empirically access the correct spectral statistics to predict the optimal schedule shape to achieve better performance. 

% \section*{Software and Data}

% We utilized the CasADi software package for numerical constrained optimal control problems \cite{andersson2019casadi}, for details see Appendix \ref{app:optimal_control}. The implementation is available in the supplementary material. We used the CIFAR-5M dataset for our ResNet experiments \cite{nakkiran2020deep}. 
% %comment out when resubmitting
% %Code to reproduce the experiments can be found at \url{https://github.com/blakebordelon/opt-control-sgd-rf}.

% % Acknowledgements should only appear in the accepted version.
% % \section*{Acknowledgements}
% % The authors thank Alex Atanasov, Francesca Mignacco, and Clarissa Lauditi for useful feedback on the manuscript. This work was supported by the Center of Mathematical Sciences and Applications (CMSA) at Harvard University.

% \section*{Impact Statement}

\paragraph{Broader Impacts} This paper presents work whose goal is to advance the field of Machine Learning. There are many potential societal consequences of our work, none of which we feel must be specifically highlighted here.

\section*{Acknowledgements}
The authors thank Alex Atanasov, Francesca Mignacco, and Clarissa Lauditi for useful feedback on the manuscript. This work was supported by the Center of Mathematical Sciences and Applications (CMSA) at Harvard University. B.B. acknowledges the Texas Advanced Computing Center (TACC) at The University of Texas at Austin for providing resources that have contributed to the research results reported within this paper, specifically allocations DMS26007 and DMS26010 on the Lonestar6-GPU system.

\bibliography{Bibliography}
\bibliographystyle{plainnat}

%%%%%%%%%%%%%%%%%%%%%%%%%%%%%%%%%%%%%%%%%%%%%%%%%%%%%%%%%%%%%%%%%%%%%%%%%%%%%%%
%%%%%%%%%%%%%%%%%%%%%%%%%%%%%%%%%%%%%%%%%%%%%%%%%%%%%%%%%%%%%%%%%%%%%%%%%%%%%%%
% APPENDIX
%%%%%%%%%%%%%%%%%%%%%%%%%%%%%%%%%%%%%%%%%%%%%%%%%%%%%%%%%%%%%%%%%%%%%%%%%%%%%%%
%%%%%%%%%%%%%%%%%%%%%%%%%%%%%%%%%%%%%%%%%%%%%%%%%%%%%%%%%%%%%%%%%%%%%%%%%%%%%%%
\newpage
\appendix
\onecolumn

\section{Long-time dynamics}
\label{app:long_time_dynamics}

This appendix derives the closed-form expression for the test loss used in the optimal-control analysis of the main text. 

\paragraph{Assumptions.} The starting point is the exact recursion for $c_{t,k}$ given in Eq.~\eqref{eq:evolution_ck}. Throughout this appendix we use:
\begin{itemize}
    \item[\textbf{(A1)}] \textbf{Gaussian features.} The features $\bm\psi(\bm x)$ are zero-mean Gaussian. Eq.~\eqref{eq:evolution_ck} follows rigorously under this assumption \cite{bordelon2022learning} and extends to non-Gaussian features under mild moment conditions.
    \item[\textbf{(A2)}] \textbf{Power-law spectra.} $\lambda_k=k^{-b}$ and $(w^*_k)^2\lambda_k=k^{-a}$, with $a,b>1$.
    \item[\textbf{(A3)}] \textbf{Zero initialization.} $\bm w_0=\bm 0$, hence $c_{0,k}=(w^*_k)^2$.
    \item[\textbf{(A4)}] \textbf{Late-time / small-eigenvalue regime.} The horizon $T$ is large and we restrict to times $t$ for which the loss is dominated by modes with $\eta_t\lambda_k\ll 1$.
    \item[\textbf{(A5)}] \textbf{Bounded label noise.} $\sigma_0$ is bounded away from zero, so that $\sigma^2$ defined in Eq.~\eqref{eq:LT_theorem1} stays of order one as $T\to\infty$.
        \item[\textbf{(A6)}] \textbf{Stability bound.} $\eta(t)\le\eta_{\max}$, where $\eta_{\max}=\mathcal{O}(\lambda_1^{-1})$ is the leading-mode stability bound of the exact recursion in Eq.~\eqref{eq:evolution_ck} and is independent of $a$ and $b$. This bound is automatic when optimizing the exact dynamics; it must be reintroduced explicitly here because (A4) drops the leading-$\lambda_k$ modes that would otherwise enforce it.

\end{itemize}

From Eq.~\eqref{eq:evolution_ck},
\begin{equation}
    c_{t+1,k}=\left(1-2\eta_t\lambda_k+\eta_t^2\frac{m_t+1}{m_t}\lambda_k^2\right)c_{t,k}+\frac{\eta_t^2}{m_t}\lambda_k\sum_{\ell=1}^{N}\lambda_\ell c_{t,\ell}+\frac{\eta_t^2}{m_t}\sigma^2\lambda_k\,.
\end{equation}
Under (A4), the term $\eta_t^2\lambda_k^2\,c_{t,k}$ is subleading compared to $\eta_t\lambda_k\,c_{t,k}$ and is dropped. Using the identity $\sum_\ell\lambda_\ell c_{t,\ell}=L_t-\sigma^2$ to absorb the second and third terms into a single $L_t$, we obtain the linear recursion
\begin{equation}
    c_{t+1,k}\approx\left(1-2\eta_t\lambda_k\right)c_{t,k}+\frac{\eta_t^2}{m_t}\lambda_k\,L_t\,.\label{eq:linearized_ck}
\end{equation}

Equation~\eqref{eq:linearized_ck} is a linear recursion in $c_{t,k}$ with $t$-dependent coefficients. Its solution reads \cite{bordelon2022learning}
\begin{equation}
    c_{T,k}=c_{0,k}\prod_{i=0}^{T-1}(1-2\eta_i\lambda_k)+\lambda_k\sum_{n=0}^{T-1}\frac{\eta_n^2}{m_n}L_n\prod_{i=n+1}^{T-1}(1-2\eta_i\lambda_k)\,.
\end{equation}

Under (A4), $\log(1-2\eta_i\lambda_k)\approx -2\eta_i\lambda_k$, so that
\begin{equation}
    \prod_{i=n+1}^{T-1}(1-2\eta_i\lambda_k)\approx \exp\!\left(-2\lambda_k\sum_{i=n+1}^{T-1}\eta_i\right)\approx e^{-2\lambda_k\chi(t_n)}\,,
\end{equation}
where we have introduced the integrated learning rate
\begin{equation}
    \chi(t):=\int_t^T\eta(t')\,dt'\,,\qquad \chi_0:=\chi(0)\,.
\end{equation}
Replacing the sum over $n$ by an integral and using (A3) to set $c_{0,k}=(w^*_k)^2$,
\begin{equation}
    c_{T,k}\approx (w^*_k)^2\,e^{-2\lambda_k\chi_0}+\lambda_k\int_0^T dt\;\frac{\eta(t)^2}{m(t)}\,L_t\,e^{-2\lambda_k\chi(t)}\,.
\end{equation}

Define
\begin{equation}
    f_s(\chi):=\sum_{k=1}^N k^{-s}e^{-k^{-b}\chi}\,.\label{eq:fs_def}
\end{equation}
Using (A2), $\sum_k\lambda_k(w^*_k)^2 e^{-2\lambda_k\chi_0}=f_a(2\chi_0)$ and $\sum_k\lambda_k^2 e^{-2\lambda_k\chi(t)}=f_{2b}(2\chi(t))$. Therefore
\begin{equation}
    L_T-\sigma^2=\sum_k\lambda_k c_{T,k}\approx f_a(2\chi_0)+\int_0^T dt\;\frac{\eta(t)^2}{m(t)}\,L_t\,f_{2b}(2\chi(t))\,.\label{eq:LT_implicit}
\end{equation}
The function $f_s$ satisfies $df_s/d\chi=-f_{s+b}(\chi)$ and $f_s(0)=\sum_{k=1}^N k^{-s}\xrightarrow[N\to\infty]{}\zeta(s)$, with the large-argument asymptotic
\begin{equation}
    f_s(\chi)\approx\frac{\Gamma((s-1)/b)}{b}\,\chi^{(1-s)/b}\,,\qquad \chi\to\infty\,.\label{eq:fs_props}
\end{equation}

Equation~\eqref{eq:LT_implicit} is implicit in $L_t$. Two observations close it. First, under (A5) the excess loss decays towards zero while $\sigma^2$ stays of order one, so $L_t-\sigma^2\ll\sigma^2$ at late times. Second, the integrand in Eq.~\eqref{eq:LT_implicit} carries the factor $f_{2b}(2\chi(t))$, which is monotonically decreasing in $\chi$ and hence concentrates the integral on times near $t=T$ where $\chi(t)$ is small; this is exactly the regime in which $L_t\approx\sigma^2$ holds with the smallest error. Substituting $L_t\to\sigma^2$ inside the integral yields the form used in the rest of the paper:
\begin{equation}
    L_T-\sigma^2\approx f_a(2\chi_0)+\sigma^2\int_0^T dt\;\frac{\eta(t)^2}{m(t)}\,f_{2b}(2\chi(t))\,.\label{eq:LT_general}
\end{equation}

\section{Analytical derivation of optimal learning-rate schedules}

\label{app:OC_analytics}

\subsection{Power-law spectra}

\label{app:OC_power_law}

For constant batch size $m(t)=m$ and the closure of Eq.~\eqref{eq:LT_general}, the optimal-control problem reduces to finding a non-increasing trajectory $\chi(t)$ with $\chi(T)=0$ that minimizes
\begin{equation}
    L_T-\sigma^2\approx f_a(2\chi_0)+\int_0^T dt\;\mathcal{L}(\dot\chi(t),\chi(t))\,,\qquad \mathcal{L}(\dot\chi,\chi)=\dot\chi^2\,A(\chi)\,,\qquad A(\chi):=\frac{\sigma^2}{m}f_{2b}(2\chi)\,,\label{eq:lagrangian_form}
\end{equation}
where we used $\eta(t)=-\dot\chi(t)$.

Since $\mathcal{L}$ has no explicit $t$ dependence, Noether's Theorem for time translations gives the conservation law $\frac{d}{dt}\!\left[\dot\chi\,\partial_{\dot\chi}\mathcal{L}-\mathcal{L}\right]=0$, which yields
\begin{equation}
    \dot\chi\,\partial_{\dot\chi}\mathcal{L}-\mathcal{L}=\dot\chi^2\,A(\chi)=c
\end{equation}
for some positive constant $c$. Equivalently, since $\dot\chi<0$,
\begin{equation}
    \dot\chi(t)=-\sqrt{\frac{c}{A(\chi(t))}}\,.\label{eq:relation_chidot}
\end{equation}

Define
\begin{equation}
    F(\chi):=\int_0^\chi d\chi'\;\sqrt{A(\chi')}\,.
\end{equation}
Separating variables in Eq.~\eqref{eq:relation_chidot} and integrating from $t$ to $T$ with $\chi(T)=0$ gives $F(\chi(t))=\sqrt{c}\,(T-t)$. At $t=0$ this fixes $\sqrt{c}=F(\chi_0)/T$, hence
\begin{equation}
    \frac{F(\chi(t))}{F(\chi_0)}=1-\frac{t}{T}\,.\label{eq:F_t_chi}
\end{equation}

Using the conservation law, the variance integral is
\begin{equation}
    \int_0^T dt\;\dot\chi(t)^2\,A(\chi(t))=c\,T=\frac{F(\chi_0)^2}{T}\,.
\end{equation}
Therefore Eq.~\eqref{eq:lagrangian_form} along the optimal trajectory reduces to a function of $\chi_0$ alone,
\begin{equation}
    L_T-\sigma^2\approx f_a(2\chi_0)+\frac{F(\chi_0)^2}{T}\,,
\end{equation}
and the stationarity condition $\partial_{\chi_0}[L_T-\sigma^2]=0$ gives
\begin{equation}
    -2\,f_{a+b}(2\chi_0)+\frac{2\,F(\chi_0)\,F'(\chi_0)}{T}=0\,,\label{eq:chi0}
\end{equation}
where we used $df_a/d\chi=-f_{a+b}$ and $F'(\chi)=\sqrt{A(\chi)}$.

For large $\chi$, the asymptotic in Eq.~\eqref{eq:fs_props} gives
\begin{equation}
    A(\chi)\approx\frac{\sigma^2}{m}\frac{\Gamma(2-1/b)}{b}\,(2\chi)^{(1-2b)/b}\,,
\end{equation}
and substituting $u=2\chi'$ in $F(\chi)=\int_0^\chi d\chi'\sqrt{A(\chi')}$, together with $\int_0^X u^{(1-2b)/(2b)}du=2b\,X^{1/(2b)}$, yields
\begin{equation}
    F(\chi)\approx\sqrt{\frac{b\,\sigma^2\Gamma(2-1/b)}{m}}\,(2\chi)^{1/(2b)}\,,\qquad F'(\chi_0)=\sqrt{A(\chi_0)}\approx\sqrt{\frac{\sigma^2\Gamma(2-1/b)}{b\,m}}\,(2\chi_0)^{(1-2b)/(2b)}\,.
\end{equation}
The product $F(\chi_0)\,F'(\chi_0)\approx (\sigma^2/m)\,\Gamma(2-1/b)\,(2\chi_0)^{(1-b)/b}$, while $f_{a+b}(2\chi_0)\approx \Gamma((a+b-1)/b)(2\chi_0)^{(1-a-b)/b}/b$. Substituting into Eq.~\eqref{eq:chi0},
\begin{equation}
    \frac{\Gamma((a+b-1)/b)}{b}\,(2\chi_0)^{(1-a-b)/b}=\frac{\sigma^2\,\Gamma(2-1/b)}{m\,T}\,(2\chi_0)^{(1-b)/b}\,.
\end{equation}
Solving for $\chi_0$ (the exponent of $2\chi_0$ on the left minus that on the right is $-a/b$),
\begin{equation}
    \chi_0\approx\frac{1}{2}\left[\frac{m}{b\,\sigma^2}\,\frac{\Gamma((a+b-1)/b)}{\Gamma(2-1/b)}\right]^{b/a}T^{b/a}\,.\label{eq:chi0_final}
\end{equation}
For $b<a$ this gives $\chi_0\sim T^{b/a}$ with $b/a<1$, consistent with $\eta(0)\sim\chi_0/T\to 0$ as $T\to\infty$. For $b>a$, $\chi_0$ grows superlinearly in $T$ and $\eta(0)$ would diverge; the constraint $\eta(t)\le\eta_{\max}$ then becomes active and is treated below.

\paragraph{Easy phase.}In the scaling limit $T\to\infty$ with $t/T$ fixed, both $\chi(t)$ and $\chi_0-\chi(t)$ are large and the asymptotic form of $A$ above is valid. Substituting into Eq.~\eqref{eq:relation_chidot},
\begin{equation}
    \dot\chi(t)\propto -\chi(t)^{1-1/(2b)}\,.
\end{equation}
Separating variables and imposing $\chi(T)=0$,
\begin{equation}
    \chi(t)\approx \chi_0\,(1-t/T)^{2b}\,,\qquad \eta(t)=-\dot\chi(t)\approx\frac{2b\,\chi_0}{T}\,(1-t/T)^{2b-1}\,,
\end{equation}
which combined with Eq.~\eqref{eq:chi0_final} reproduces the closed-form schedule of the main text. Substituting Eq.~\eqref{eq:chi0_final} into $L_T-\sigma^2\approx f_a(2\chi_0)+F(\chi_0)^2/T$ and using the asymptotic forms above, both terms scale as $T^{-(a-1)/a}$, yielding the easy-phase loss scaling.

\paragraph{Hard phase.}In the hard phase, the constraint $\eta\leq \eta_{\max}$ becomes active and we find the optimal schedule 
\begin{equation}
    \eta(t)=\begin{cases}
        \eta_{\max} \quad &\text{ for }t<t_s\,,\\
        \eta_{\max}\left(\frac{1-t/T}{1-t_s/T}\right)^{2b-1}\quad &\text{ for }t>t_s\,,
    \end{cases}\label{eq:lr_hard}
\end{equation}
where the switching time $0<t_s<T$ can be determined by minimizing the generalization error. It is useful to define $\tilde{\eta}(\tau)=\eta(\tau T)$. Then, the loss can be written as
\begin{equation}
  L_T-\sigma^2\sim T^{-(a-1)/b}\left[ \int_0^1 d\tau \ \tilde{\eta}(\tau) \right]^{-(a-1)/b} + \frac{\sigma^2}{m}  T^{-(b-1)/b} \int_0^1 d\tau \ \tilde{\eta}(\tau)^2  \left[ \int_{\tau}^1 d\tau'   \ \tilde{\eta}(\tau') \right]^{- \frac{2b-1}{b}}\,.
\end{equation}
If $b>a$, the first term will dominate, implying $t_s\approx T$ to leading order. If $b<a$, the second term will dominate. As shown below, the second term is minimized by $t_s=0$, therefore the constraint $\eta<\eta_{\max}$ is never saturated for $b<a$, in agreement with the easy-phase analysis above. We therefore focus on the case $b>a$ and compute the first-order corrections to $\tau_s\coloneqq t_s/T\approx 1$.

\begin{figure}
    \centering
    \includegraphics[width=0.5\linewidth]{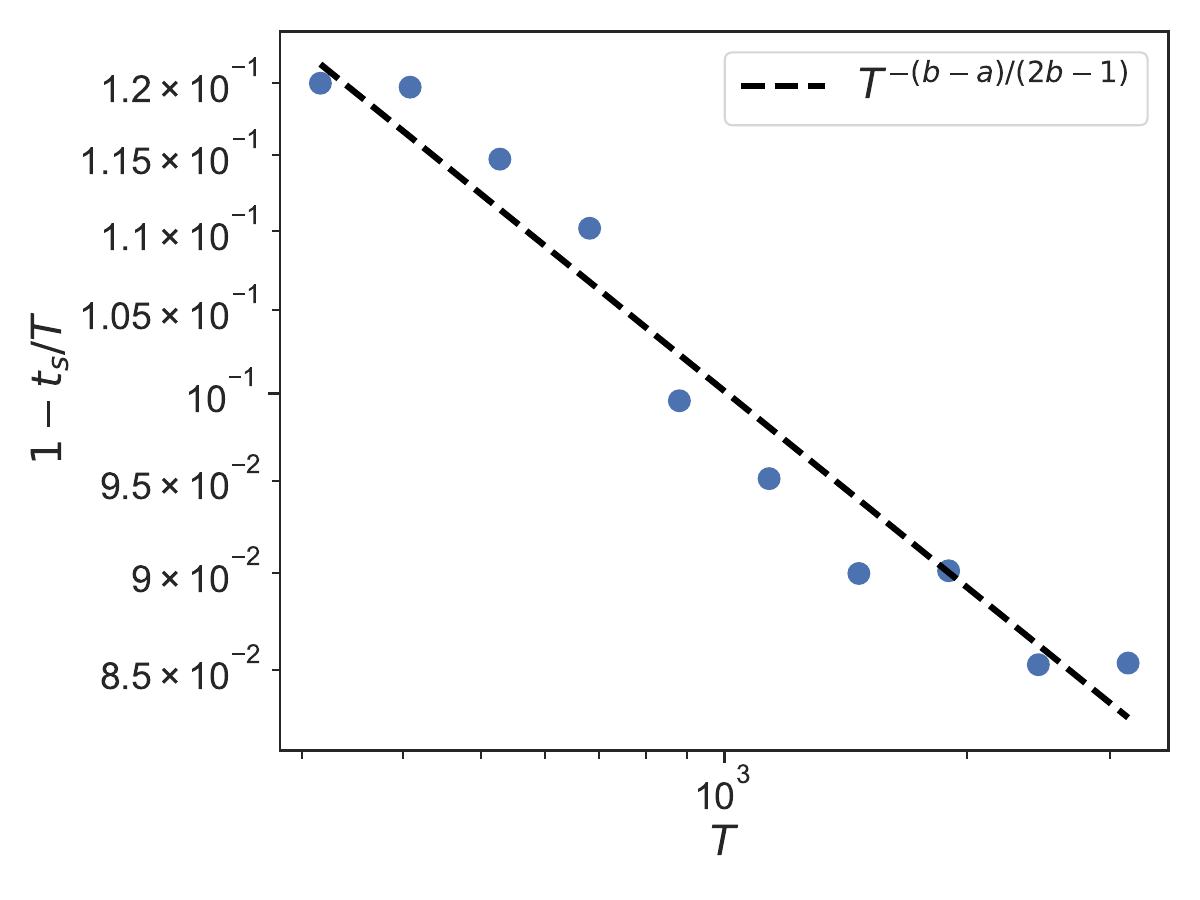}
    \caption{Fraction $1-t_s/T$ of the total training time spent in the annealing regime. The switching time $t_s$ is defined here as the time at which the schedule $\eta^*(t)$ crosses the level $\eta=0.15$ for the first time. Same parameters as in Fig.~\ref{fig:combined_results}a.}
    \label{fig:ts}
\end{figure}

Using Eq.~\eqref{eq:lr_hard}, we find
\begin{align}
  L_T-\sigma^2\sim T^{-(a-1)/b} f_{\rm Bias}(\tau_s) + \frac{\sigma^2}{m} T^{-(b-1)/b}f_{\rm Var}(\tau_s) \,,
\end{align}
where 
\begin{equation}
  f_{\rm Bias}(\tau_s)=  \left[ \tau_s+\frac{1-\tau_s}{2b} \right]^{-(a-1)/b}\,,
\end{equation}
and
\begin{equation}
   f_{\rm Var}(\tau_s)= \left[ \frac{b}{b-1}\left(\left(\frac{1-\tau_s}{2b}\right)^{-(b-1)/b}-\left(\tau_s+\frac{1-\tau_s}{2b}\right)^{-(b-1)/b}\right) +(2b)^{(2b-1)/b}(1-\tau_s)^{-(b-1)/b}\right]
\end{equation}
Note that the variance second term is an increasing function of $\tau_s$, confirming that in the easy phase $b<a$, the constraint $\eta(t)<\eta_{\max}$ is never saturated. In the hard case $b>a$, we consider corrections $\tau_s\approx 1-\epsilon$ and expand to leading order for small $\epsilon$, yielding (up to constants)
\begin{align}
    L_T-\sigma^2\sim T^{-(a-1)/b}+ T^{-(a-1)/b} \epsilon+  T^{-(b-1)/b}\epsilon^{-(b-1)/b} \,.
\end{align}
Minimizing this expression over $\epsilon$, we identify the scaling
\begin{equation}
    \epsilon\sim T^{-(b-a)/(2b-1)}\,.
\end{equation}
Overall, this argument predicts that
\begin{equation}
    t_s\approx T(1- c T^{-(b-a)/(2b-1)})
\end{equation}
for some constant $c$. Figure \ref{fig:ts} shows that this prediction agrees with the result obtained by optimizing the full dynamics with CasADi (in that case, for a schedule $\eta(t)$ we define $t_s$ as the first time at which $\eta(t)=0.15$). As a result, the loss scales as
\begin{align}
    L_T-\sigma^2 \sim T^{-(a-1)/b}+\sigma^2 T^{-(b-1)/b}\epsilon^{-(b-1)/b} \sim T^{-(a-1)/b} +\sigma^2 T^{-(b-1)(a+b-1)/(b(2b-1))}\,.\label{eq:LT_hard}
\end{align}

\begin{figure*}[htbp]
    \centering
    
    \begin{subfigure}[b]{0.45\linewidth}
        \centering
        \includegraphics[width=\linewidth]{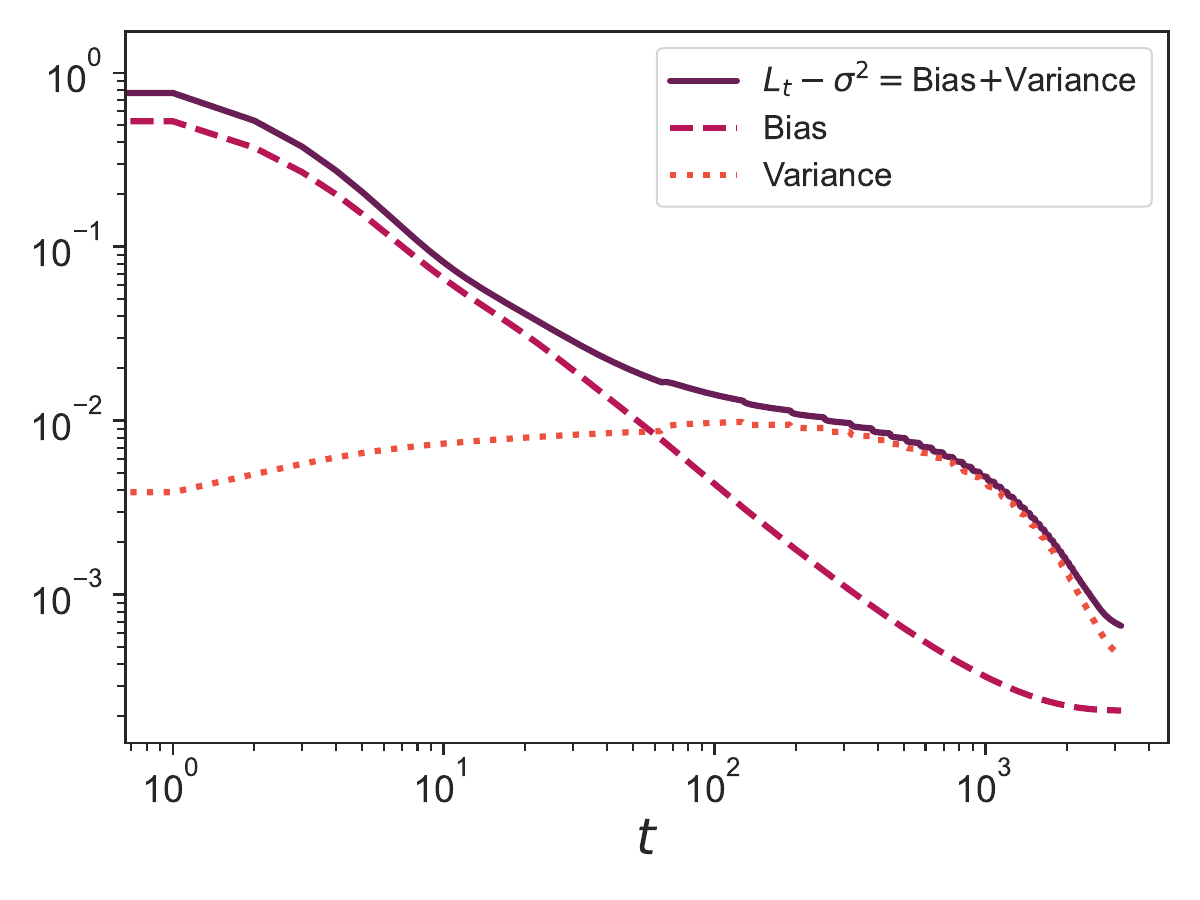}
        \caption{Easy Phase ($b< a$)}
       
    \end{subfigure}
    \hfill
    \begin{subfigure}[b]{0.45\linewidth}
        \centering
        \includegraphics[width=\linewidth]{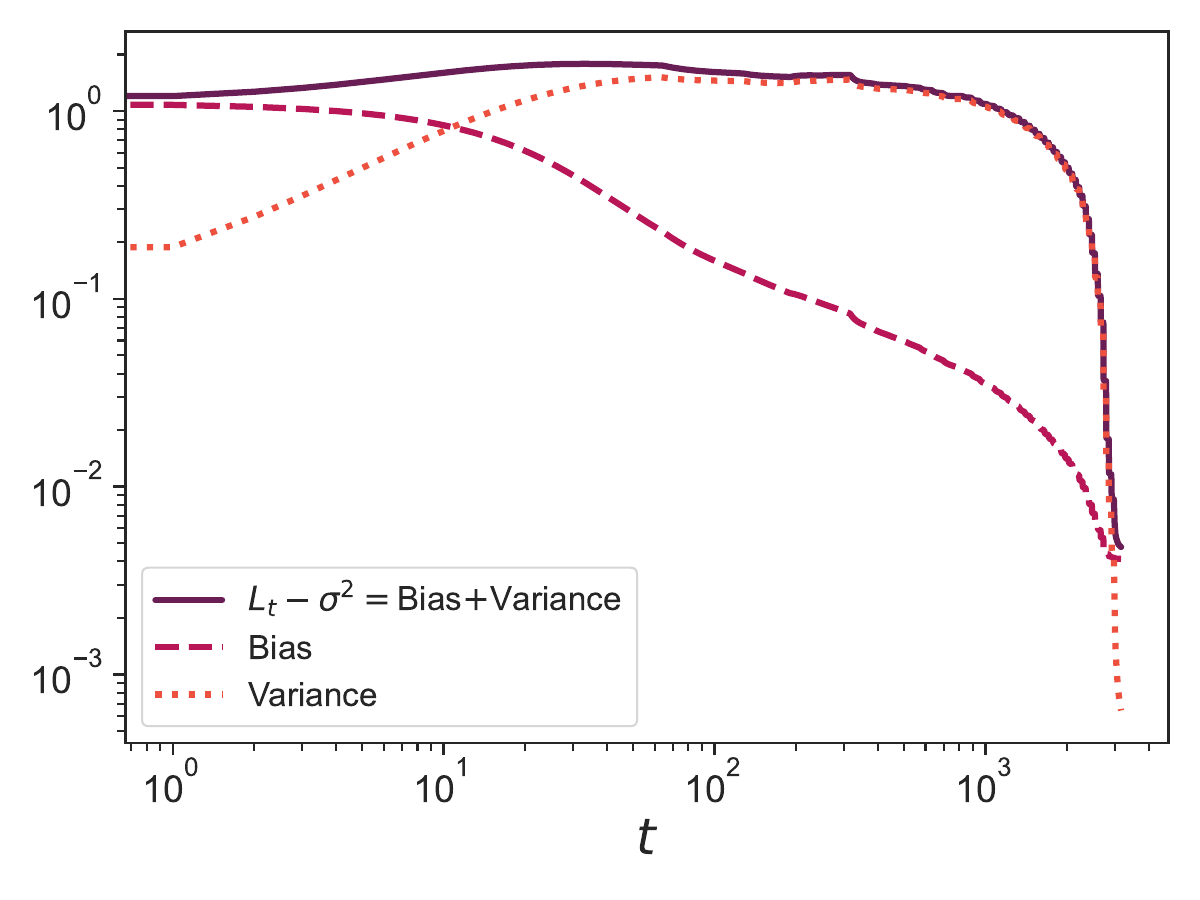}
        \caption{Hard Phase ($b> a$)}
    \end{subfigure}

    \caption{\textbf{Decomposition of the excess loss $L_t - \sigma^2$ into bias and variance components.}  \textbf{(a)} In the easy phase ($b=2$, $a=3.5$), the optimal schedule minimizes bias and variance simultaneously throughout the training trajectory. 
    \textbf{(b)} In the hard phase ($b=5$, $a=3.5$), the schedule minimizes the bias for the majority of the training time ($t < t_s$) where the learning rate is large, while the final annealing phase ($t > t_s$) is responsible for suppressing the variance.
    \textbf{Parameters:} $T=3162$, $N=1000$, $\sigma=0.5$, $m=5$. }
    \label{fig:loss_decomposition}
\end{figure*}

\subsection{Compute-optimal scaling}
\label{app:compute_optimal}

\begin{figure*}[htbp]
    \centering
    
    \begin{subfigure}[b]{0.45\linewidth}
        \centering
        \includegraphics[width=\linewidth]{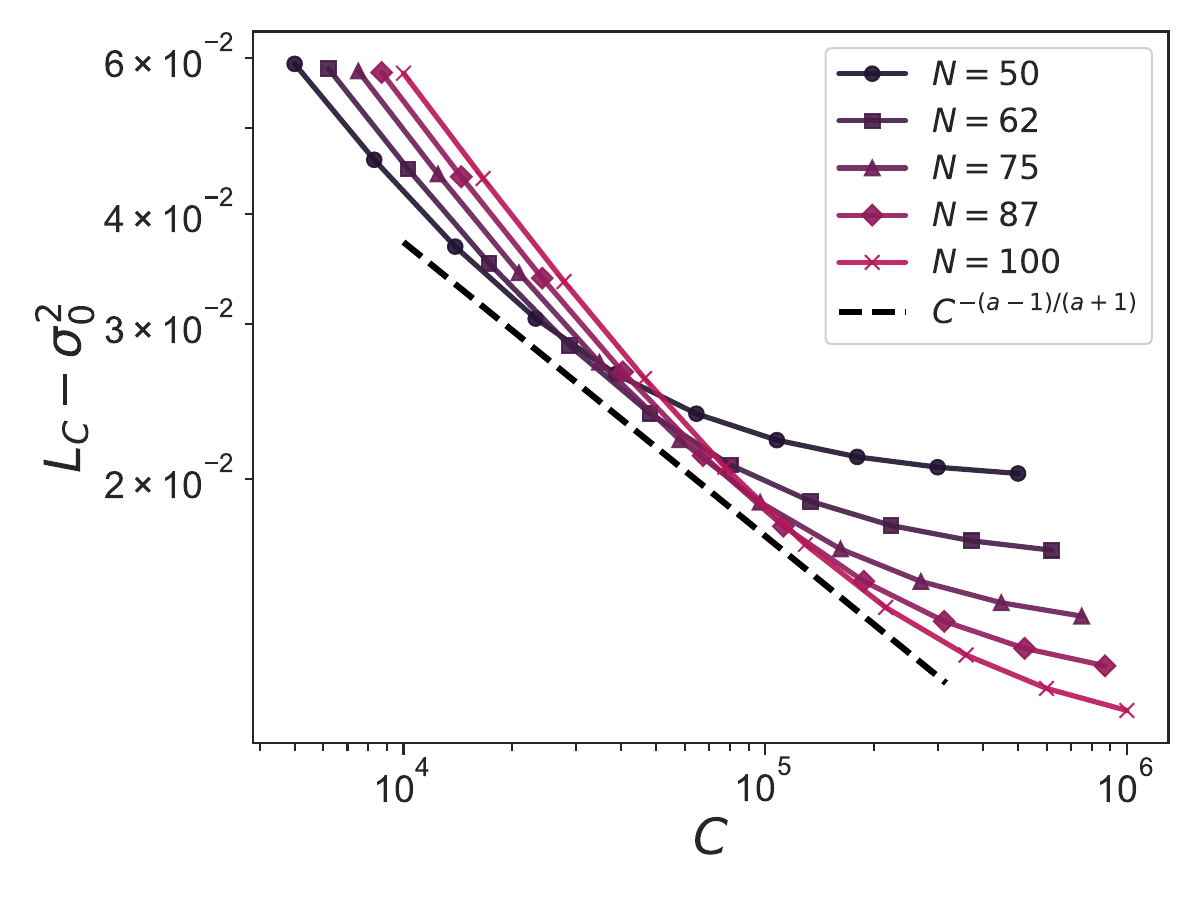}
        \caption{Easy Phase ($b< a$)}
       
    \end{subfigure}
    \hfill
    \begin{subfigure}[b]{0.45\linewidth}
        \centering
        \includegraphics[width=\linewidth]{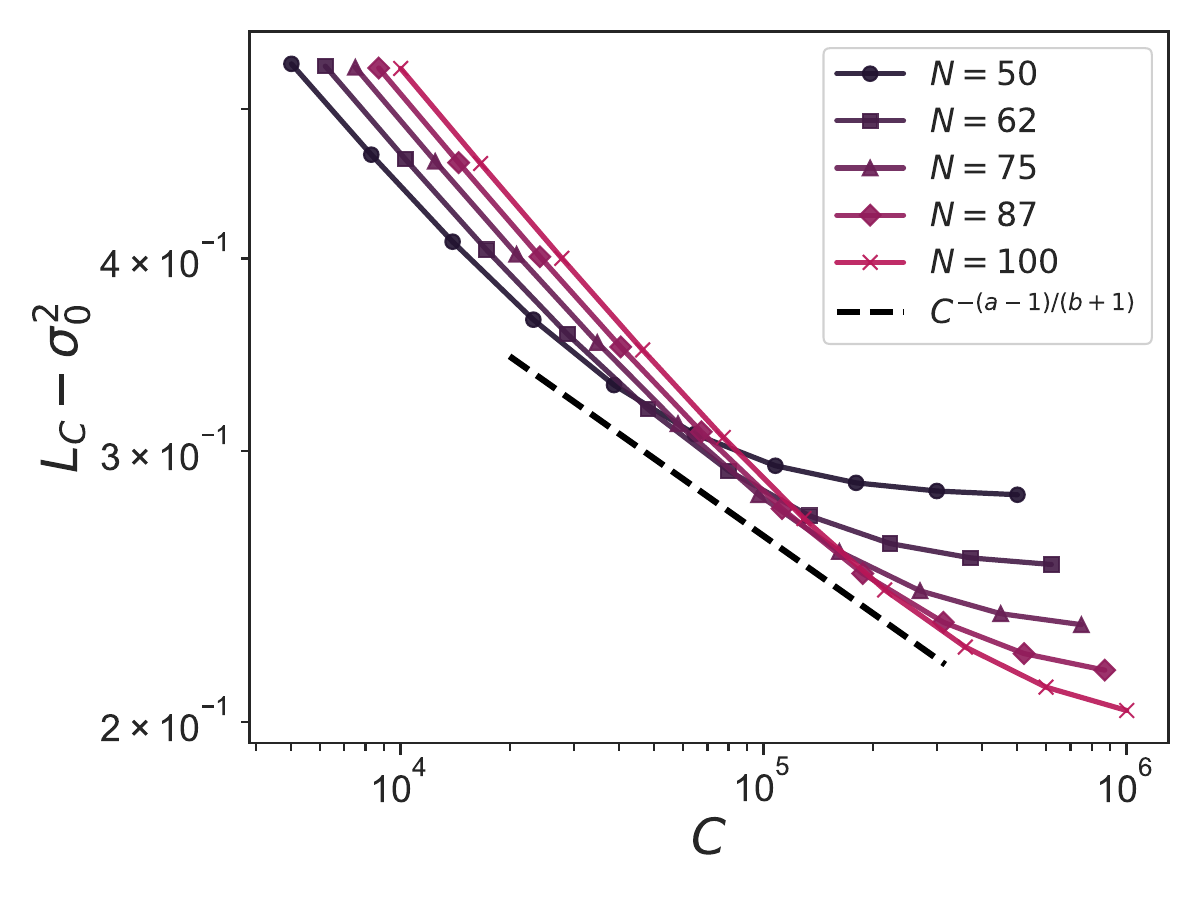}
        \caption{Hard Phase ($b> a$)}
    \end{subfigure}

    \caption{\textbf{Compute optimal scaling.}  Residual loss $L_C-\sigma_0^2$ as a function of the compute $C=mNT$ for different values of the model size $N$. The dashed lines indicate the theoretical prediction. \textbf{Parameters:} $\sigma=0.5$, $m=5$. In the easy phase $b=1.5$ and $a=2$, in the hard phase $b=2$ and $a=1.5$. }
    \label{fig:scaling}
\end{figure*}

We next address the problem of jointly scaling the model size $N$ and the data budget $B_{\rm tot}$ to minimize the generalization error $L_C$ under a fixed compute budget $C = B_{\rm tot}N$. Using Result \ref{thm:opt_LR_powerlaw}, for $M \gg N \gg 1$, the effective noise variance scales as $\sigma^2 \sim \sigma_0^2 + N^{1-a}$. Consequently, in the easy phase, the excess loss scales as $L_T - \sigma_0^2 \sim B_{\rm tot}^{-(a-1)/a} + N^{1-a}$. Balancing these terms yields the compute-optimal allocation $B_{\rm tot} \sim C^{a/(a+1)}$ and $N \sim C^{1/(a+1)}$. This results in an optimal error scaling of $L_C - \sigma_0^2 \sim C^{-(a-1)/(a+1)}$. Similarly, in the hard phase, the loss scales as $L_T - \sigma_0^2 \sim B_{\rm tot}^{-(a-1)/b} + N^{1-a}$. Optimizing this tradeoff leads to $B_{\rm tot} \sim C^{b/(b+1)}$ and $N \sim C^{1/(b+1)}$, yielding a final error scaling of $L_C - \sigma_0^2 \sim C^{-(a-1)/(b+1)}$. These theoretical scaling laws are verified in Fig.~\ref{fig:scaling}.

\section{Comparison with benchmarks}

\label{app:benchmarks}

\subsection{Constant learning rate}

For a constant learning rate $\eta_T(t)=\eta_T$ we have $\chi(t)=\eta_T(T-t)$, with $\chi_0=\eta_T T$. Substituting into Eq.~\eqref{eq:LT_general} (with $m(t)=m$),
\begin{equation}
    L_T-\sigma^2\approx f_a(2\eta_T T)+\frac{\sigma^2\,\eta_T^2}{m}\int_0^T dt\;f_{2b}(2\eta_T(T-t))\,.
\end{equation}
Changing variables $s=T-t$ and then $\chi'=2\eta_T s$,
\begin{equation}
    \int_0^T dt\;f_{2b}(2\eta_T(T-t))=\int_0^T ds\;f_{2b}(2\eta_T s)=\frac{1}{2\eta_T}\int_0^{2\eta_T T}d\chi'\;f_{2b}(\chi')\,.
\end{equation}
The integral evaluates via the antiderivative $df_b/d\chi=-f_{2b}$ together with $f_b(0)=\zeta(b)$ (large $N$),
\begin{equation}
    \int_0^{2\eta_T T}d\chi'\;f_{2b}(\chi')=\zeta(b)-f_b(2\eta_T T)\;\xrightarrow[2\eta_T T\to\infty]{}\;\zeta(b)\,,
\end{equation}
since $f_b(2\eta_T T)\sim (2\eta_T T)^{(1-b)/b}\to 0$ for $b>1$. The bias term satisfies $f_a(2\eta_T T)\approx \Gamma((a-1)/b)(2\eta_T T)^{(1-a)/b}/b$. Combining,
\begin{equation}
    L_T-\sigma^2\approx \frac{\Gamma((a-1)/b)}{b}\,(2\eta_T T)^{-(a-1)/b}+\frac{\zeta(b)\,\sigma^2\,\eta_T}{2\,m}\,.
\end{equation}
Dropping $T$-independent prefactors,
\begin{equation}
    L_T - \sigma^2 \sim (\eta_T T)^{-\frac{a-1}{b}} + \frac{\sigma^2 \eta_T}{m} \,.
\end{equation}
Optimizing over $\eta_T$, we find
$\eta_T\sim  T^{-(a-1)/(a+b-1)}$ and $L_T-\sigma^2\sim T^{-(a-1)/(a+b-1)}$.

\subsection{Power-law schedules}

In this section, we investigate the optimal power-law schedule $\eta(t)\sim t^{-\delta}$, with $0<\delta<1$. We start from the expression
\begin{equation}
    L_T-\sigma^2
    \sim \left[ \int_0^T ds \ \eta(s) \right]^{-\alpha} + \frac{\sigma^2}{m} \int_0^T dt \ \eta(t)^2  f_{2b}(\chi(t))=\text{Bias}(T)+\text{Variance}(T)\,,\label{eq:loss_bias_variance}
\end{equation}
where $\alpha=(a-1)/b$ and
\begin{equation}
    \text{Bias}(T)=\left[ \int_0^T ds \ \eta(s) \right]^{-(a-1)/b}\sim T^{-\alpha(1-\delta)}\,.
\end{equation}
The variance integral can be decomposed in two contributions: one from the ``bulk'' of the integral and one from the region close to the upper limit $T$. 
\begin{equation}
    \text{Variance}(T)=\int_0^T dt \ \eta(t)^2  f_{2b}(\chi(t))\approx\int_0^{T-\tau} dt \ \eta(t)^2  f_{2b}(\chi(t))+\int_{T-\tau}^T dt \ \eta(t)^2  f_{2b}(\chi(t))\,,
\end{equation}
where we choose $\tau$ such that $\chi(T-\tau)\sim \mathcal{O}(1)$, leading
\begin{equation}
   \chi(T-\tau)\coloneqq \int_{T-\tau}^T \eta(t')dt'\sim  T^{1-\delta}-(T-\tau)^{1-\delta}\approx \tau T^{-\delta}\,,
\end{equation}
so that $\tau\sim T^{\delta}$. As a consequence, for $t<T-\tau$, we can approximate $f_{2b}(\chi(t))\approx \chi(t)^{-\gamma}\sim T^{-\gamma(1-\delta)}$ with $\gamma=(2b-1)/b$, while for $t>T-\tau$, $f_{2b}(\chi(t))$ is approximately constant. Therefore,
\begin{equation}
    \text{Variance}(T)\sim T^{-\gamma(1-\delta)} \int_0^{T-\tau} dt \ t^{-2\delta} +\int_{T-\tau}^T dt \ T^{-2\delta}= T^{-\gamma(1-\delta)} \frac{T^{1-2\delta}-1}{1-2\delta}+ T^{-\delta}\,.
\end{equation}
Therefore, $\delta$ must be chosen to minimize the combination of four terms
\begin{equation}
   L_T-\sigma^2 \sim T^{-\alpha(1-\delta)}+T^{-\gamma (1-\delta)}+T^{-\gamma (1-\delta)+1-2\delta}+T^{-\delta}\,,
\end{equation}
where we recall that $\alpha=(a-1)/b>0$ and $1<\gamma=2-1/b<2$. We find
\begin{equation}
\delta=\min\left[ \frac{\alpha}{\alpha+1},\frac{\gamma}{\gamma+1} \right]
\end{equation}
with a crossover between the two regimes at $a=2b$ and 
\begin{equation}
   L_T-\sigma^2\sim T^{-\delta}\,.
\end{equation}
Note that for the optimal schedule
\begin{equation}
 L_T-\sigma^2\sim T^{-\min((a-1)/a,(a-1)/b)}
\end{equation}
so that there is an exponent gap between the optimal and the optimal power law schedules for all $a,b>1$.

We next consider the scaling of power-law schedules with $T$-dependent prefactors of the form $\eta(t)\sim T^{-\nu}t^{-\delta}$. Following the same steps as before, we find
\begin{equation}
 L_T-\sigma^2\sim T^{\alpha\nu-\alpha(1-\delta)}+T^{-2\nu-\gamma (1-\delta-\nu)}+T^{-2\nu-\gamma (1-\delta-\nu)+1-2\delta}+T^{-\delta-\nu}\,.
\end{equation}
Denoting $x=\delta+\nu$, we can rewrite 
\begin{equation}
     L_T-\sigma^2\sim T^{-\alpha(1-x)}+T^{-(\gamma+(2-\gamma)x-2\delta)}+T^{-\gamma+1-(2-\gamma)x}+T^{-x}\,.
\end{equation} 
Therefore, we can choose $\delta=0$, leading to the same scaling behavior as the constant-in-$t$ optimal-in-$T$ schedule.

\subsection{General scaling form}

In this section, we investigate the performance of scaling form schedules
\begin{equation}
    \eta(t)= T^{-\xi}g(t/T)\,,\label{eq:eta_scaling_form}
\end{equation}
for $\xi\geq 0$. Note that in the easy phase $b<a$, the optimal schedule corresponds to $\xi=1-b/a$ and $g(z)\sim (1-z)^{2b-1}$. We find (for $m=1$)
\begin{equation}
 L_T-\sigma^2\sim \left[\int_0^T dt~ \eta(t)\right]^{-\alpha}+\sigma^2\int_0^T dt~\eta^2(t) f_{2b} \left(\int_t^T dt'~\eta(t')\right)\,,
\end{equation}
where $\alpha=(a-1)/b$ and $f_{2b}(\chi)= \sum_{k=1}^N k^{-2b} e^{-k^{-b}\chi}$. Plugging the expression in Eq.~\eqref{eq:eta_scaling_form} and changing variables $\tau=t/T$ and $\tau'=t'/T$, we find
\begin{equation}
    L_T-\sigma^2\approx \text{Bias}(T)+ \text{Variance}(T)\,,
\end{equation}
where
\begin{equation}
    \text{Bias}(T)=T^{-\alpha(1-\xi)}\left[\int_0^1 d\tau~ g(\tau)\right]^{-\alpha}
\end{equation}
and
\begin{equation}
    \text{Variance}(T)=\sigma^2 T^{1-2\xi}\int_0^1 d\tau~g^2(\tau) ~f_{2b} \left(T^{1-\xi}\int_{\tau}^1 d\tau'~g(\tau')\right)
\end{equation}
We recall the asymptotic behaviors $f_s(\chi)\approx \zeta(s)=\sum_{k\geq 1}k^{-s}$ for small $\chi$ (and large $N$) and $f_s(\chi)\approx \Gamma((s-1)/b) \chi^{(1-s)/b}/b$ for large $\chi$. It is then instructive to split the variance integral into an edge component, where $\chi(t)=\int_t^T dt'~\eta(t')\ll 1$ and a bulk component where $\chi(t)\gg 1$. To do so, we define $t^*$ so that $\chi(t^*)=1$, yielding the condition for $\tau^*\coloneqq t^*/T$
\begin{equation}
    \int_{\tau^*}^1d\tau' g(\tau')=T^{\xi-1}\,.
\end{equation}
Assuming $\xi<1$ (since for $\xi>1$ the bias increases with $T$), we get that $\tau^*$ is determined by the asymptotic behavior of $g(\tau)$ for $\tau\to 1$. Assuming $g(\tau)\sim (1-\tau)^{\delta}$ with $\delta>0$, we get
\begin{equation}
    (1-\tau^*)\sim T^{-(1-\xi)/(1+\delta)}\,.
\end{equation}
We can then split the variance term into a 
``bulk'' component for $\tau<\tau^*$ and an ``edge'' component for $\tau>\tau^*$, leading to
\begin{equation}
    \text{Variance}(T)\approx\text{Bulk Variance}(T)+\text{Edge Variance}(T) \,.
\end{equation}
 The bulk contribution reads
\begin{equation}
    \text{Bulk Variance}(T)\sim  T^{1-2\xi}\int_0^{\tau^*} d\tau~g^2(\tau) ~ \left(T^{1-\xi}\int_{\tau}^1 d\tau'~g(\tau')\right)^{-\gamma}\,.
\end{equation}
For $\delta>(\gamma-1)/(2-\gamma)=b-1$ the integral over $\tau$ converges as $\tau^*\to 1$ and hence
\begin{equation}
    \text{Bulk Variance}(T)\sim  T^{1-2\xi-\gamma (1-\xi)}\,.
\end{equation}
However, if $\delta<b-1$ there is an additional contribution from the divergence as $\tau^*\to 1 $ for large $T$, leading to
\begin{equation}
    \text{Bulk Variance}(T)\sim  T^{-(\delta+\xi)/(\delta+1)}\,.
\end{equation}

The edge contribution is
\begin{align}
    \text{Edge Variance}(T)&\sim  T^{1-2\xi}\int_{\tau^*}^1 d\tau~g^2(\tau) \sim T^{1-2\xi}\int_{\tau^*}^1 d\tau~(1-\tau)^{2\delta} \sim T^{1-2\xi}(1-\tau^*)^{2\delta+1}\\
    &\sim T^{1-2\xi-(1-\xi)(1+2\delta)/(1+\delta)}\sim T^{-(\delta+\xi)/(\delta+1)}\,.
\end{align}
Overall, for $\delta>b-1$, we get
\begin{equation}
      L_T-\sigma^2\sim T^{-\alpha(1-\xi)}+T^{1-2\xi-\gamma (1-\xi)}+T^{-(\delta+\xi)/(\delta+1)}\,,
\end{equation}
while for $\delta<b-1$ we find
\begin{equation}
      L_T-\sigma^2\sim T^{-\alpha(1-\xi)}+T^{-(\delta+\xi)/(\delta+1)}\,,
\end{equation}
Interestingly, for large enough $\delta$, the edge variance terms is negligible and the exponent $\xi$ is obtained by matching the bias and bulk variance terms. In particular, we find $\xi=(a-b)/a$ for $b<a$ (easy phase) and $\xi=0$ for $b>a$ (hard phase). The corresponding scaling laws are $L_T-\sigma^2\sim T^{-(a-1)/a}$ for $b<a$ and $L_T-\sigma^2\sim T^{-(a-1)/b}$ for $b>a$. The scalings match those of the optimal schedule. 

\subsubsection{The edge term is subleading for the optimal schedule}

In the easy phase $b<a$, the optimal schedule takes the form in Eq.~\eqref{eq:eta_scaling_form} with $\xi=(a-b)/a$ and $\delta=2b-1$. This value of $\delta$ is chosen to minimize the prefactor of the scaling law. For this value of $\delta$, the edge variance term, which was not included in the derivation of the optimal schedule, is indeed subleading.

We next consider the edge variance term of the optimal schedule in the hard phase $b>a$. Recall that in this phase the optimal learning rate schedule is (taking $\eta_{\max}=1$)
\begin{equation}
    \eta(t)=\begin{cases}
        1 \quad &\text{ for }t<t_s\,,\\
        \left(\frac{1-t/T}{1-t_s/T}\right)^{\delta}\quad &\text{ for }t>t_s\,,
    \end{cases}
\end{equation}
with $\delta=2b-1$ and $T-t_s\sim T^{1-(b-a)/(2b-1)}$. The edge boundary $t^*$ is determined by the condition $\chi(t^*)=1$, yielding
\begin{equation}
   \chi(t^*)= \int_{t^*}^T dt' \eta(t')= \int_{t^*}^T dt' \left(\frac{1-t'/T}{1-t_s/T}\right)^{\delta}=T \epsilon^{-\delta}(1-\tau^*)^{\delta+1}=1\,,
\end{equation}
where we have assumed $t_s<t^*$ and we have defined $\epsilon\coloneqq 1-t_s/T\sim T^{-(b-a)/(2b-1)}$ and $\tau^*=t^*/T$. Therefore, we get
\begin{equation}
    1-\tau^*=\left(\frac{\epsilon^{\delta}}{T}\right)^{1/(\delta+1)}\sim T^{-(b-a+1)/(2b)}\,.
\end{equation}
Note that $1-t_s/T>1-\tau^*$, confirming that $t_s<t^*$.

The edge variance term reads
\begin{equation}
    \text{Edge Variance}(T)=\int_{t^*}^T dt~\eta^2(t)=\int_{t^*}^T dt~\eta^2(t)= \int_{t^*}^T dt~\left(\frac{1-t/T}{1-t_s/T}\right)^{2\delta}=\left(\epsilon T\right)^{-\delta/(\delta+1)}\,.
\end{equation}
The bulk variance term scales as $(\epsilon T)^{-1+1/b}$, confirming that the edge term is subleading.

\section{Joint optimization of learning rate and batch size}
\label{app:batch}

We want to find the joint optimal schedule of learning rate and batch size that minimizes the final loss. We consider a constraint of the total number of training examples $B_{\rm tot}=\int_0^T dt ~m(t)$.

Using Eq.~\eqref{eq:LT_general}, the loss with the data-budget constraint reads
\begin{equation}
   L_T -\sigma^2 \approx f_a(2\chi_0)+\sigma^2\int_0^T dt\;\frac{\eta^2(t)}{m(t)}\,f_{2b}(2\chi(t))+\mu\int_0^T dt\;m(t)\,,
\end{equation}
where the Lagrange multiplier $\mu$ enforces the constraint on $m(t)$. Optimizing over $m(t)$ yields
\begin{equation}
m(t)=\eta(t)\,\sigma\sqrt{\frac{f_{2b}(2\chi(t))}{\mu}}\,,
\end{equation}
and the constraint $B_{\rm tot}=\int_0^T dt\,m(t)$ then fixes
\begin{equation}
    \sqrt{\mu}=\frac{\sigma}{B_{\rm tot}}\int_0^{\chi_0}d\chi\;\sqrt{f_{2b}(2\chi)}\,,
\end{equation}
where we changed integration variable using $\eta(t)\,dt=-d\chi$. Plugging back into the loss,
\begin{equation}
    L_T-\sigma^2 \approx f_a(2\chi_0)+\frac{\sigma^2}{B_{\rm tot}}\left[\int_0^{\chi_0}d\chi\;\sqrt{f_{2b}(2\chi)}\right]^2\,.
\end{equation}
Note that this expression depends on $\chi_0$ only, so at fixed $\chi_0$ all learning-rate schedules give the same performance provided $m(t)$ is chosen accordingly.
\begin{figure}
    \centering
    \includegraphics[width=\linewidth]{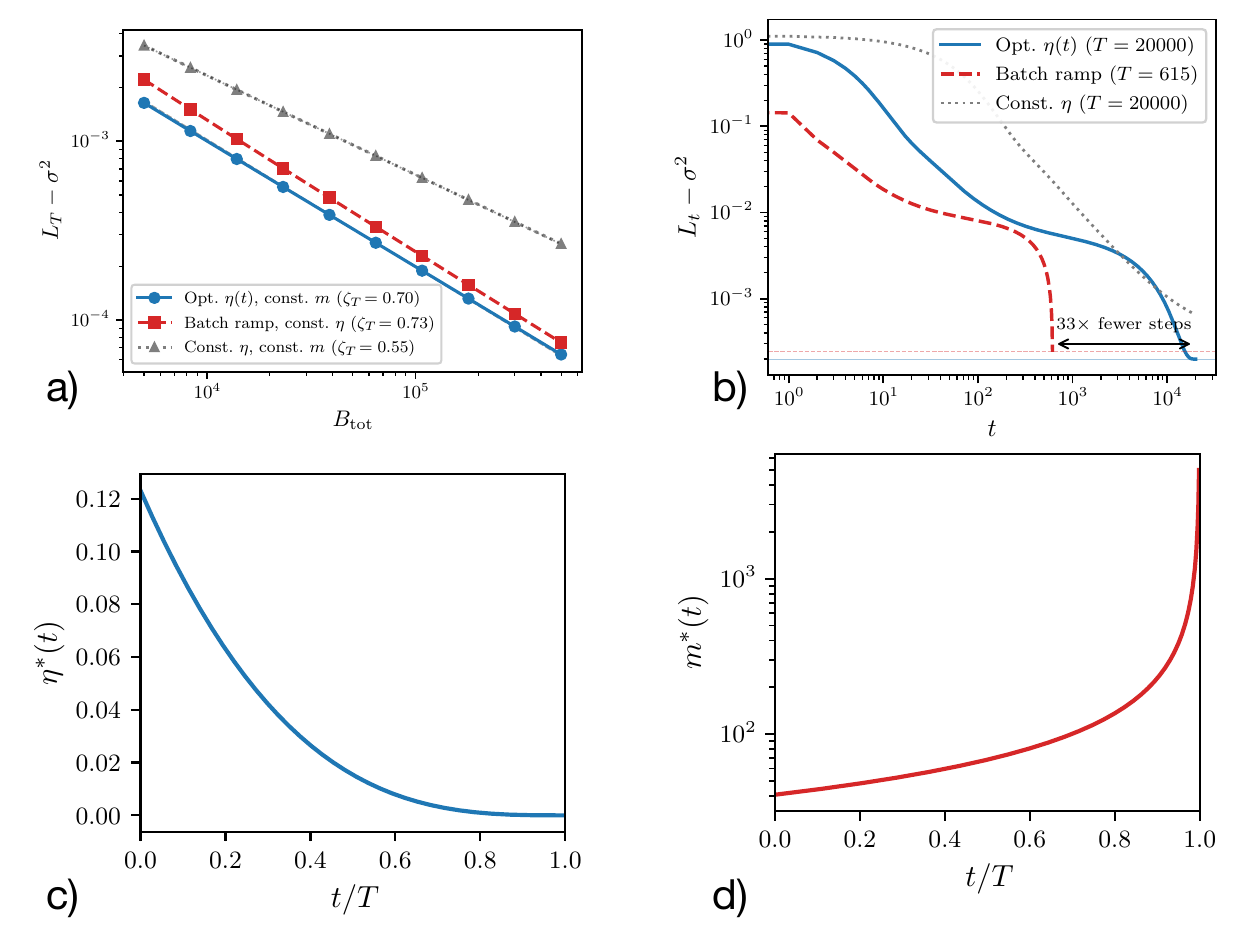}
    \caption{{\bf Optimal batch schedule in the easy phase ($b<a$).} We simulate the deterministic loss dynamics (Eq.~\eqref{eq:evolution_ck}) of the powerlaw random feature model and compare three strategies at fixed total data budget $B_{\rm tot}$: the optimal learning rate schedule (panel c) with constant batch size $m$, the batch ramp (panel d) with constant $\eta=\eta_{\max}$, and a constant learning rate baseline (with $\eta$ chosen optimally at each $B_{\rm tot}$). {\bf a)} Final excess loss $L_T-\sigma^2$ as a function of $B_{\rm tot}$. Both the optimal schedule and the batch ramp achieve the same scaling exponent $\zeta_T\approx (a-1)/a\approx 0.71$, outperforming the constant baseline ($\zeta_T\approx (a-1)/(a+b-1)\approx 0.56$). {\bf b)} Loss dynamics for $B_{\rm tot}=10^5$. The batch ramp reaches the same loss scaling in $33\times $ fewer wall-clock steps, confirming that the joint optimization reduces wall-clock time while preserving the optimal scaling with respect to $B_{\rm tot}$. {\bf Parameters:} $N=1000$, $a=3.5$, $b=2$, $\sigma=0.5$, $m=5$, $\eta_{\max}=1$.}
    \label{fig:batch_ramp}
\end{figure}

For large $\chi_0$, we find 
\begin{equation}
    L_T-\sigma^2\sim \chi_0^{-(a-1)/b}+\chi_0^{1/b}/B_{\rm tot}\,,
\end{equation}
implying that at optimality $\chi_0\sim B_{\rm tot}^{b/a}$. This scaling can be achieved in the easy phase ($b<a$), leading to $ L_T-\sigma^2\sim B_{\rm tot}^{-(a-1)/a}$. Note however that $\chi_0$ cannot scale superlinearly with $B_{\rm tot}$. Indeed, since $\eta(t)\leq \eta_{\max}$ and $m(t)\geq 1$, we have the conditions $\chi_0\leq \eta_{\rm max}T$ and $B_{\rm tot}\geq T$, yielding $\chi_0\leq \eta_{\max} B_{\rm tot}$. Therefore, in the hard phase, the optimum is achieved for $\chi_0\sim B_{\rm tot}\sim T$, for which $L_T\sim B_{\rm tot}^{-(a-1)/b}$.

In both easy and hard phases, the degeneracy in the optimization over $\eta$ and $m$ can be broken by requiring the solution to have the minimum wall-clock time $T$ (see Fig.~\ref{fig:batch_ramp}). This corresponds to taking $\eta(t)=\eta_{\max}$ and 
\begin{equation}
    m(t)=\frac{B_{\rm tot}}{2bT}(1-t/T)^{1/(2b)-1}\,.
\end{equation}
In the hard phase, $B_{\rm tot}\sim T$, so that $m(t)$ remains bounded. However, in the easy phase $T\sim B_{\rm tot}^{b/a}$ so that the batch size actually grows with $B_{\rm tot}$ as $m(t)\sim B_{\rm tot}^{1-b/a}$.

\section{Loss Dynamics for SGD + Momentum}\label{app:sgd_momentum_derivation}

\subsection{SGD Derivation}

In this section, we compute the average case loss dynamics over random draws of data. This follows the standard arguments in \cite{bordelon2022learning}. Starting from the following dynamics for $\bm \Delta_{t} = (\bm w^\star - \bm w_t)$. This variable has dynamics
\begin{align}
    \bm\Delta_{t+1} = \bm \Delta_{t} - \eta_t \bm g_t \ , \  \bm g_t = \frac{1}{m_t} \sum_{\mu=1}^{m_t} \tilde{\bm\psi}_{\mu,t} [ \tilde{\bm\psi}_{\mu,t} \cdot \bm \Delta_{t} + \sigma \epsilon_{\mu,t} ]
\end{align}
The correlation matrix of interest is $\bm C_t = \left< \bm \Delta_t \bm \Delta_t^\top \right>_{\mathcal D_{t-1}}$. The loss is $L_t = \text{Tr} \bm\Lambda \bm C_{t} + \sigma^2$. This matrix satisfies the exact recursion
\begin{align}
    \bm C_{t+1} = \bm C_{t} - \eta_t \bm\Lambda \bm C_t - \eta_t \bm C_t \bm\Lambda + \eta_t^2 \left< \bm g_t \bm g_t^\top \right> .
\end{align}
For Gaussian features, the gradient correlations have the following form
\begin{align}
    \left< \bm g_t \bm g_t^\top \right> = \frac{m_t+1}{m_t}  \bm\Lambda \bm C_t \bm \Lambda + \frac{1}{m_t} \bm\Lambda \left[ \text{Tr} \bm\Lambda \bm C_t + \sigma^2 \right] .
\end{align}
Since the loss only depends on the diagonal entries of $\bm C_t$ which we define as $c_{t,k} \coloneqq C_{t,kk}$.
\begin{align}
    c_{t+1,k} = \left( 1- 2 \eta_t \lambda_k + \eta_t^2  \frac{m_t+1}{m_t} \lambda_k^2 \right) c_{t,k} +  \frac{\eta_t^2}{m_t} \lambda_k \sum_{\ell} \lambda_{\ell} c_{t,\ell} + \frac{\sigma^2 \eta_t^2}{m_t} \lambda_k . 
\end{align}
Under upper and lower bounds on the fourth moments of the features, one can derive analogous upper and lower bounds for the loss dynamics and SGD noise \cite{bordelon2022learning}.

\subsection{Momentum}

To analyze momentum, we must also track auxiliary momentum variable $\bm v(t)$, we track the following updates
\begin{align}
    \bm\Delta_{t+1} = \bm\Delta_t - \eta_t \bm v_t \ , \ \bm v_t =  (1-\beta_t) \bm v_{t-1} + \beta_t \bm g_t \quad , \quad \bm g_t = \frac{1}{m} \sum_{\mu=1}^m \tilde{\bm\psi}_{\mu,t} [ \tilde{\bm\psi}_{\mu,t} \cdot \bm \Delta_{t} + \sigma \epsilon_{\mu,t} ]
\end{align}
To compute the test loss, we need to track the following three correlation functions
\begin{align}
    C_{k,t} = \left< \Delta_{k,t}^2  \right> \ , \ V_{k,t} = \left< v_{k,t}^2 \right> \ , \ R_{k,t} = \left<  \Delta_{k,t} v_{k,t} \right>
\end{align}
where the average is computed over the random samples from each minibatch. These three variables evolve according to the following linear dynamics  
\begin{align}
    &C_{k, t+1} = C_{k,t} - 2 \eta_{t} R_{k,t} + \eta_t^2 V_{k,t} \,,\nonumber
    \\
    &V_{k,t+1} = (1-\beta_t)^2 V_{k,t} + 2 \beta_t (1-\beta_t) \lambda_k ( R_{k,t} - \eta_t V_{k,t} ) \nonumber\\ &+ \beta_t^2 \left[  \frac{m+1}{m} \lambda_k^2 C_{k,t+1} + \frac{1}{m} \lambda_k \sum_{\ell} \lambda_{\ell} C_{\ell, t+1} + \frac{\sigma^2}{m} \lambda_k \right]  \nonumber
    \\
    &R_{k,t+1} = (1-\beta_t) ( R_{k,t} - \eta_t V_{k,t} ) + \beta_t \lambda_k C_{k,t+1}  
\end{align}
We note that due to the time indexing that we adopted above, the right hand side for the equations defining the dynamics for $R_{k,t+1}$ and $V_{k,t+1}$ depend on $C_{k,t+1}$ rather than $C_{k,t}$ (though of course this can be substituted).\footnote{We find this convention yields better stability than defining the dynamics as $\bm \Delta_{t+1} = \bm \Delta_{t} - \eta_t \bm v_{t}$ where $\bm v_{t} = (1-\beta_t) \bm v_{t-1} + \beta_t \bm g_{t-1}$.} The loss function dynamics, as before, can be computed from $C_{k,t}$
\begin{align}
    L_t - \sigma^2 = \sum_{k} \lambda_k C_{k,t} .
\end{align}
We use the above dynamics for $\{ C_{k,t}, R_{k,t}, V_{k,t} \}$.

\section{Optimal Control Implementation}

\label{app:optimal_control}

To determine the optimal learning rate schedules numerically, we formulate the problem as a discrete-time optimal control problem governed by the recursive evolution of $c_{t,k}$ (Eq. \eqref{eq:evolution_ck}). We employ a direct single-shooting method to transcribe this control problem into a non-linear programming (NLP) problem. In this formulation, the control inputs (the learning rate $\eta_t$ and optionally the batch size $m_t$) are treated as the primary decision variables. The state trajectory is not included in the set of decision variables. The objective function, defined as the terminal generalization error $L_T$, is thus expressed as a direct function of the initial conditions and the sequence of control inputs.

We implement this optimization framework using CasADi~\cite{andersson2019casadi}, which performs automatic differentiation to efficiently compute the necessary gradients of the objective with respect to the control trajectory. The resulting NLP is solved using IPOPT (Interior Point OPTimizer), a primal-dual interior-point method. To ensure computational efficiency for large horizons, we configure the solver to use a limited-memory approximation of the Hessian and the MUMPS linear solver for the internal step computations. The implementation of our optimization framework is available in the supplementary material.

\section{Experiment and Compute Details}

\begin{figure}
    \centering
\includegraphics[width=0.4\linewidth]{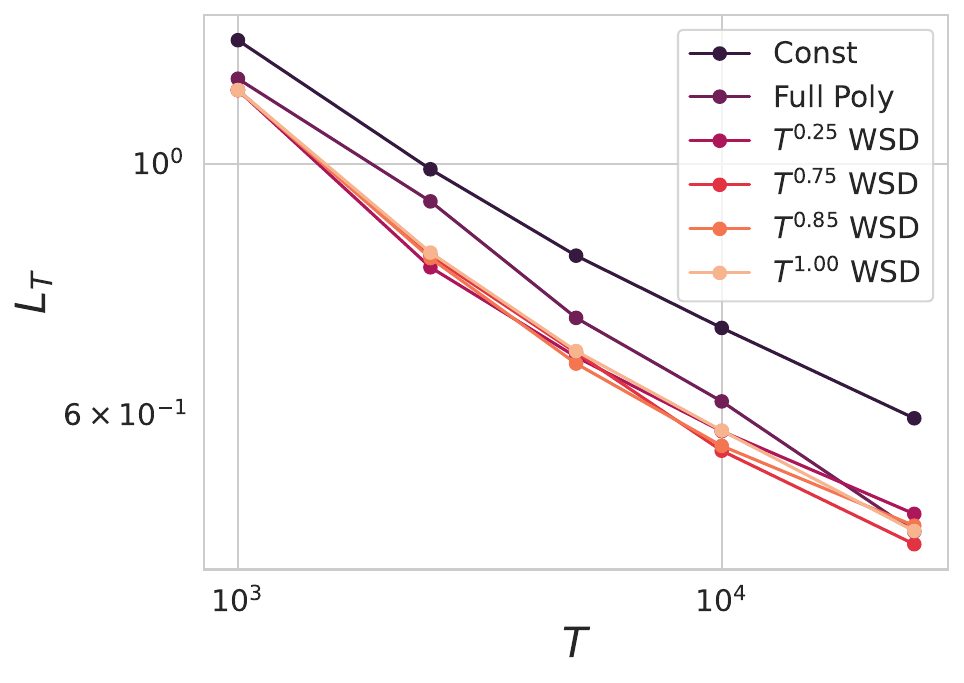}
    \caption{ResNet CIFAR-5M training contrasting annealing for $T^\delta$ steps instead of $O(T)$ steps. As predicted by our theory in the hard phase, these schedules outperform fully polynomial annealing.  }
    \label{fig:wsd_compare_cifar}
\end{figure}

\paragraph{ResNets on CIFAR-5M} We train deep residual convolutional networks on CIFAR-5M with SGD and different schedules. We contrast different WSD like schedules predicted by our hard phase optimal annealing strategies in Figure \ref{fig:wsd_compare_cifar}. Experiments were conducted on a Nvidia A100 GPU and took about 10-15 minutes per run. 

\paragraph{Language Model} The models are based on the OLMo repository models and training scripts \cite{olmo20242olmo2furious}. The base model is $D_{\text{model}} = 768$ with $H=12$ heads and $L=12$ blocks attention and MLP blocks. The C4 data are tokenized with the GPT-2 tokenizer whose vocabulary is $\approx 50$k. Weight decay is $0.1$ and either a constant or cosine annealing schedule is adopted.  We used 3x Nvidia A100 per run for varying token horizon $T$. The runtime ranges between 1-10 hours depending on $T$. 

\newpage

%%% Add checklist back for full sub
%\input{checklist.tex}

\end{document}